\documentclass[12pt]{article}
\usepackage{a4}
\usepackage{epsf}
\usepackage{bbold}

\begin{document}

\def\bea{\begin{eqnarray}}
\def\eea{\end{eqnarray}}
\def\be{\begin{equation}}
\def\ee{\end{equation}}
\def\ss{\scriptstyle}
\def\Tr{\mbox{Tr}}
\newcommand\rf[1]{(\ref{#1})}
\def\nn{\nonumber}
\newcommand{\sect}[1]{\setcounter{equation}{0} \section{#1}}
\renewcommand{\theequation}{\thesection .\arabic{equation}}
\newcommand{\NPB}[3]{{\it Nucl.\ Phys.} {\bf B#1} (#2) #3}
\newcommand{\CMP}[3]{{\it Commun.\ Math.\ Phys.} {\bf #1} (#2) #3}
\newcommand{\PRD}[3]{{\it Phys.\ Rev.} {\bf D#1} (#2) #3}
\newcommand{\PLB}[3]{{\it Phys.\ Lett.} {\bf B#1} (#2) #3}
\newcommand{\JHEP}[3]{{JHEP} {\bf #1} (#2) #3}
\newcommand{\hepth}[1]{{\tt hep-th/#1}}
\newcommand\eqn[1]{eq.~(\ref{#1})}
\newcommand{\binomial}[2]{\left ( \matrix{#1\cr #2\cr}\right )}
\newcommand\cO{{\cal O}}
\newcommand{\ft}[2]{{\textstyle\frac{#1}{#2}}}
\def\st{\scriptstyle}
\def\sst{\scriptscriptstyle}
\def\ra{\rightarrow}
\def\lra{\longrightarrow}
\newcommand\Zb{\bar Z}

\thispagestyle{empty}
\begin{flushright}
{\sc\footnotesize hep-th/0205033}\\
{\sc AEI 2002-036}
\end{flushright}
\vspace{1cm}
\setcounter{footnote}{0}
\begin{center}
{\Large{\bf A New Double-Scaling Limit of ${\cal N}=4$ Super
Yang-Mills Theory and PP-Wave Strings}
    }\\[14mm]
{\sc C. Kristjansen\footnote{Permanent Address: The 
Niels Bohr Institute, Blegdamsvej 17, Copenhagen \O, DK2100 Denmark.  
Work supported by the Danish Natural Science Research Council.},  
J. Plefka, G. W. Semenoff\footnote{Work supported 
in part by NSERC of Canada.  
Permanent Address: Department of Physics and Astronomy, 
University of British Columbia, Vancouver, British Columbia  
V6T 1Z1.} and M. Staudacher\\[7mm]
Max-Planck-Institut f\"ur Gravitationsphysik\\
Albert-Einstein-Institut\\
Am M\"uhlenberg 1, D-14476 Golm, Germany}\\ [2mm]
{\tt kristjan,plefka,semenoff,matthias@aei.mpg.de}\\[20mm]

{\sc Abstract}\\[2mm]
\end{center}
The metric of a spacetime with a parallel plane (pp)-wave can be
obtained in a certain limit of the space AdS$^5\times $S$^5$.
According to the AdS/CFT correspondence, the holographic dual of
superstring theory on that background should be the analogous limit of
${\cal N}=4$ supersymmetric Yang-Mills theory.  
In this paper we shall show that, contrary to widespread expectation, 
non-planar diagrams survive this limiting procedure in the gauge theory.
Using matrix model techniques as well as combinatorial reasoning
it is demonstrated that a subset of diagrams of arbitrary genus
survives and that a non-trivial {\it double scaling limit} may be
defined. We exactly compute two- and three-point functions
of chiral primaries in this limit. We also carefully study 
certain operators conjectured to correspond to string excitations 
on the pp-wave background. We find non-planar linear mixing
of these proposed operators, requiring their redefinition. 
Finally, we show that the redefined operators receive 
non-planar corrections to the planar one-loop anomalous dimension.

%We compute some protected Yang-Mills theory correlators in this limit
%and obtain and interpret the corresponding scaling functions. 
%These results are complemented by the calculation of the leading contributions
%to (unprotected) momentum correlators as well as their one-loop
%quantum corrections. Finally we discuss matching of the double scaling
%limit with the summation over all genera in string theory on the 
%pp-wave background. \vfill
\leftline{{\sc May 2002}}

\newpage
\setcounter{page}{1}

\sect{Introduction and Conclusions}

The AdS/CFT correspondence asserts a duality between type IIB
superstring theory quantized on the background space
AdS$^5\times$S$^5$ and ${\cal N}=4$ supersymmetric Yang-Mills theory
in four dimensional Minkowski space.  Recently, it has been observed
that a certain limit of the AdS$^{5}\times$S$^{5}$ geometry yields a
parallel plane-wave (pp-wave) space-time \cite{Blau:2002dy}.  This
background has the virtue that the superstring can readily be
quantized there \cite{Metsaev:2001bj}. The pp-wave geometry is
obtained in a large radius of curvature and large angular momentum
limit. In the Yang-Mills theory dual, which was discussed in
\cite{Berenstein:2002jq}, this translates to the limit
\begin{equation}\label{limit}
N \to \infty ~~{\rm and}~~ J\to\infty~~{\rm with} ~~
\frac{J^2}{N}~~{\rm fixed},~~g_{{\rm{{\rm YM}}}}~~{\rm fixed}
\end{equation}
where $N$ is the rank of the U(N) gauge group and $J$ is the isospin
quantum number which is conjugate to the phase of the complex
combination of two scalar fields
$Z(x)=\frac{1}{\sqrt{2}}\left(\Phi_1(x)+i\Phi_2(x)\right)$.  The
Yang-Mills coupling constant $g_{{\rm YM}}$ (and the string coupling
$g_s=g_{{\rm YM}}^2/4\pi$) are held constant and a new parameter, $J^2/N$
appears in the limit. The validity of such large J limits has
been discussed in \cite{Polyakov:2001af}.
The BPS bound $\Delta \geq J$ implies that the
operators of interest have large conformal dimensions $\Delta$.  The
simplest example of such an operator is the chiral primary field 
\be
{1 \over \sqrt{J} N^{J/2}}~\Tr~Z^J(x)
\label{loop}
\ee
which saturates the bound.
More generally, one is interested in correlation functions containing a large
number (of ${\cal O}(J)$) $Z(x) Z(x) \ldots Z(x)$ of fields
such that $\Delta - J$ stays finite as $J \rightarrow \infty$. 
Note that for all $J$ the fields (\ref{loop}) 
are protected operators: Their two and three point functions do not 
receive quantum corrections beyond the free field sector of super
Yang-Mills theory. 
  
In \cite{Berenstein:2002jq} it has been assumed that 
the gauge theory remains planar in the limit (\ref{limit}). This assumption,
which was used extensively in \cite{Berenstein:2002jq},
deserves to be studied more carefully. 
In fact\footnote{{\em Cf} the discussion of this
point in a footnote on page 8 of \cite{Berenstein:2002jq}.
Actually, in \cite{Balasubramanian:2001nh}
it was shown that if one scales $J \sim N$ non-planar diagrams
{\it dominate} over planar ones. This is consistent with our finding
that the scaling $J \sim \sqrt{N}$ corresponds precisely
to the critical situation where (the generic) non-planar diagrams are
neither dominant nor subdominant w.r.t.~planar diagrams.} it 
had been observed in
\cite{Balasubramanian:2001nh}
that, if $J$ tends to infinity sufficiently rapidly with $N$, non-planar
diagrams will eventually dominate over planar ones. 
In this paper we shall point out that the particular scaling
given in eq.(\ref{limit}) corresponds to the most interesting case 
where some (but not all) of the non-planar diagrams are surviving
the limiting procedure. This allows us to demonstrate that 
eq.(\ref{limit}) actually corresponds to replacing the standard
't Hooft limit by a an interesting novel {\it double-scaling limit} 
of Yang-Mills theory. It will be shown below that in some ways this 
new limit resembles the double scaling limit of the 
``old matrix models'' of non-critical bosonic string theory discovered
in \cite{Brezin:rb}. One consequence is that, in order to keep
exact orthonormality of the operators (\ref{loop}), their correct
normalization involves a non-trivial scaling function
\be
\Bigg( J N^J
\frac{
\sinh\left( \frac{1}{2}\frac{J^2}{N}\right) }{ \frac{1}{2}
\frac{J^2}{N} } \Bigg)^{-\frac{1}{2}}
~\Tr~Z^J(x)
\label{newloop}
\ee

After establishing that the recipe (\ref{limit}) does not fully
suppress non-planar diagrams we are immediately led to the following
puzzling question: What, then, characterizes the classes of diagrams
that are favored, respectively suppressed, in this novel limit? By
carefully investigating the pertinent combinatorics we find the
following picture: Interpreting the operator $\Tr Z^J $ as a discrete
closed string consisting of $J$ ``string bits'' (see
\cite{Berenstein:2002jq} and references therein) non-planar diagrams
contributing to this operator correspond to the string splitting into
multiple strings in intermediate channels. Taking $J$ large may be
interpreted as a continuum limit: The number of string bits diverges
and the string becomes long and macroscopic. On the other hand, taking
$N$ large acts towards suppressing the string splitting.  We then find
that the scaling $J^2 \sim N$ leads to a delicate balance between the
two effects such that microscopic strings, made out of only a small
number of string bits (small w.r.t.~$J$), are suppressed but
macroscopic strings, made out of a large number of bits (i.e.~of
${\cal O}(J)$), survive.

We also look at the three point correlation function of the chiral
primaries (\ref{newloop}), which are protected as well, and find
the exact scaling function. As it explicitly encodes information for 
arbitrary genus, it would be fascinating if its
structure could be understood from the 
string side \cite{Spradlin:2002ar}.

The present picture is very clear in the case of operators which are
protected by supersymmetry, such as (\ref{newloop}).  In the case of
unprotected operators, their quantum corrections require further
analysis.  Interactions involve index loops which produce factors of
$N$.  In the 't Hooft limit of Yang-Mills theory, these factors are
controlled by making the coupling constant small, $g^2_{{\rm YM}}\sim 1/N$.
In the limit (\ref{limit}), $g_{{\rm YM}}$ is finite, so we depend on the
large $J$ limit to suppress factors of $N$.  We shall present
evidence, based on a one-loop computation,  that this
suppression indeed occurs in two-point correlators of
the operators\footnote{As we shall see below
the sum in eq.(\ref{momop}) should actually start at $l=0$.}

\be 
\cO_n(x) = \frac{1}{\sqrt{J}\,
N^{J/2+1}}\, \sum_{l=1}^J \Tr [ \, \phi_3 \, Z^l\, \phi_4\,
Z^{J-l}\, ]\, e^{\frac{2\pi\, i \, n\, l}{J}}
\label{momop}
\ee 
which were introduced in ref.\cite{Berenstein:2002jq}.  
There, it was argued that these operators correspond to states of the
string theory on the pp-wave background, 
$
\cO_n \leftrightarrow a_{n}^{\dagger\,8}  a_{-n}^{7} |0,p_+\rangle
$, 
and that their conformal
dimensions should match eigenvalues of the pp-wave string Hamiltonian, 
\begin{equation}\label{spectrum}
E =  \sum_{n=0}^\infty N_n \sqrt{ 1+ \frac{ 4\pi g_s N}{J^2}n^2 }
\end{equation} 
where $N_n$ is the occupation number of a state of excitation level
$n$.  As a simple check of the duality, ref.\cite{Berenstein:2002jq}
showed that this formula is indeed reproduced by the planar limit of
Yang-Mills theory. In this paper, as a warm-up exercise, 
we will reproduce this computation in detail.  

The precise definition of the operators (\ref{momop}) is actually
quite subtle. As for (\ref{newloop}), non-planar diagrams also survive
even at the classical level. 
But unlike (\ref{newloop}), here it is not sufficient
to replace the planar normalization factor 
$(J N^{J+2})^{-\frac{1}{2}}$ by a scaling function since even
their orthogonality is violated at the non-planar level.  
Furthermore, at one-loop order, non-planar diagrams survive the limit
(\ref{limit}) as well and contribute to the two-point correlator of
(\ref{momop}). The main effect of the non-planar diagrams then is to
mix these operators so that linear combinations of them are required
to obtain {\it redefined} operators with a fixed conformal dimension. 
One of our central results is that after this redefinition, 
at the one-loop level and to first order in ${J^4 \over N^2}$, 
the limit (\ref{limit}) is well-defined for these operators.

Finally, we find that there is a non-planar contribution to the
scaling dimension of the redefined operators\footnote{
In the first version of this paper we had overlooked this contribution
due to some erroneous analyticity assumptions in the evaluation of
a minor subset of the
required sums in the scaling limit. After that, the paper 
\cite{Constable:2002hw} appeared, which contained the correct treatment of
this correction. In addition, \cite{Constable:2002hw} 
interprets, under some assumptions, this result using string field theory,
while otherwise nicely
confirming our results concerning the existence of the double scaling
limit as well as the phenomenon of operator mixing.}: 
eq.(\ref{spectrum}) receives corrections of order 
$g^2_{{\rm YM}}\frac{J^2}{N}$. Relying on the proposed
duality \cite{Berenstein:2002jq}, this implies that the pp-wave string
spectrum is renormalized by string loops.
It would be very interesting to extend this result to higher
orders in $g^2_{\rm YM}$ and $\frac{J^4}{N^2}$.

\sect{Double scaling limit of chiral primary two-point functions} 
The simplest example of
double-scaling occurs in the normalization of the chiral primary
operator $ {\rm Tr} \left( Z^J(x) \right) $.  This operator saturates
a BPS bound, $\Delta-J=0$ and in the limit (\ref{limit}) it is
identified with the BPS ground state of the string theory sigma model
on the pp-wave background.  Its normalization can be computed from its
two-point function,
\begin{equation}\label{norm}
G^{JJ'}(x)=
\langle 0| {\rm Tr} \left(  Z^J(x) \right)
{\rm Tr} \left(  \bar Z^{J'}(0) \right)
|0\rangle
\end{equation}
U(1) symmetry implies that this vanishes unless $J=J'$.  
Non-renormalization theorems exist for the two- and three-point
functions of chiral primaries. These correlation functions
are then given exactly by their 
free field limits. In particular, this means that the scaling
dimension of such operators is equal to its free field limit.  
Four-point functions of these operators are known to receive 
radiative corrections. 

%Then, to proceed further, we must recall some facts about 
%chiral primary operators.
%Chiral primary operators are formed by taking the symmetric, 
%traceless components  of products of the scalar fields
%$$
%{\cal O}^{(k)}(x)=\frac{1}{N}
%{\rm Tr} \left( \Phi^{i_1}(x)\Phi^{i_2}(x)\ldots\Phi^{i_k}(x)\right)
%$$ 
%Here $i_j=1,\ldots,6$ transforms under the SO(6) R-symmetry group.  
%These operators are special in the sense that they commute with half
%of the supercharges and therefore transform as members of a short 
%multiplet of the supersymmetry algebra.  
%This fact can be used to prove no-renormalization theorems for their
%two and three-point functions - which 

Thus, we can evaluate (\ref{norm}) in free field theory. There, conformal
invariance gives the exact coordinate dependence of the correlator and
it remains to solve the combinatorial problem of taking into account
all contractions of the free field propagators.  The latter is
conveniently summarized by a correlator in a Gaussian complex matrix
model\footnote{Note that we use the standard normalization for a
complex matrix model (see appendix A). Since our calculations in the
field theory are done using $\Tr (T_a T_b)=\frac{1}{2} \delta_{a,b}$
Wick contractions of the {\it field} $Z(x)$ come with an additional
factor of $\frac{1}{2}$ compared to the {\it matrix Z}.}, 
\begin{equation}
G^{JJ'}(x)= 
\frac{\delta^{JJ'}}{|x|^{2J}} \left(\frac{g_{{\rm YM}}^2}{8\pi^2}\right)^J
\langle {\rm Tr}Z^J~{\rm Tr}\bar Z^J\rangle
\label{mmmodel}
\end{equation}
where
\begin{equation}
\langle {\rm Tr}Z^J~ {\rm Tr}\bar Z^J \rangle = 
\int dZ d\bar Z ~{\rm Tr} Z^J ~ {\rm Tr} \bar Z^J 
~e^{-{\rm Tr}\bar Z Z }  
\label{mmodel}
\end{equation}
Hereafter, 
in this paper, we shall use the notation that $\langle 0|...|0\rangle $
corresponds to the vacuum expectation value in the quantum field
theory defined by the action (\ref{symaction}) and $\langle ...\rangle$ 
denotes an expectation value of the appropriate matrices in the matrix model
defined by (\ref{mmodel}).

The correlator (\ref{mmodel}) can be computed using matrix model
techniques (see Appendix A).  The result is 
\begin{equation}
\langle {\rm Tr}Z^J~ {\rm Tr}\bar Z^J \rangle = \frac{1}{J+1}\left(
\frac{\Gamma(N+J+1)}{\Gamma(N)} - \frac{\Gamma(N+1)}{\Gamma(N-J)}
\right)
\label{corr}
\end{equation}
where it is assumed that $N>J>0$.  

The result eq.(\ref{corr}) is simple and explicit and will enable
us to understand the nature of the scaling limit eq.(\ref{limit}).
In fact it is straightforward to expand eq.(\ref{corr}) as
a series in ${1 \over N^2}$ and extract, for general $J$, the
corrections to the (trivial) planar limit $J N^J$
\be
\langle {\rm Tr}Z^J~ {\rm Tr}\bar Z^J \rangle=
{2 N^J \over J+1} \sum_{h=0}^{\infty}
\frac{1}{N^{2h}}
\left\{
\sum_{1\leq   i_1 <i_2 < \ldots i_{2h+1} \leq J}
i_1\cdot i_2\cdot \ldots \cdot i_{2h+1} \right\}
\ee
The terms in this expansion can be organized in an interesting way as
follows
\begin{eqnarray}
\langle {\rm Tr}Z^J~ {\rm Tr}\bar Z^J \rangle & =&
J~N^J  \Bigg\{ 1 + \Bigg[ {J \choose 4} + {J \choose 3}\Bigg]~{1
  \over N^2} \cr 
& & + \Bigg[ 21 {J \choose 8}+49 {J \choose 7}
+36 {J \choose 6} + 8 {J\choose 5} 
\Bigg] {1 \over N^4} + \ldots 
\Bigg\} \cr
& = & J N^J \Bigg\{ 1+ \sum_{h=1}^{\infty} \sum_{k=2 h+1}^{4 h}
a_{h,k} {J \choose k} {1 \over N^{2 h}} \Bigg\}
\label{expansion}
\end{eqnarray}

\begin{figure}[t]
\centerline{{\epsfxsize=4.5cm\epsfbox{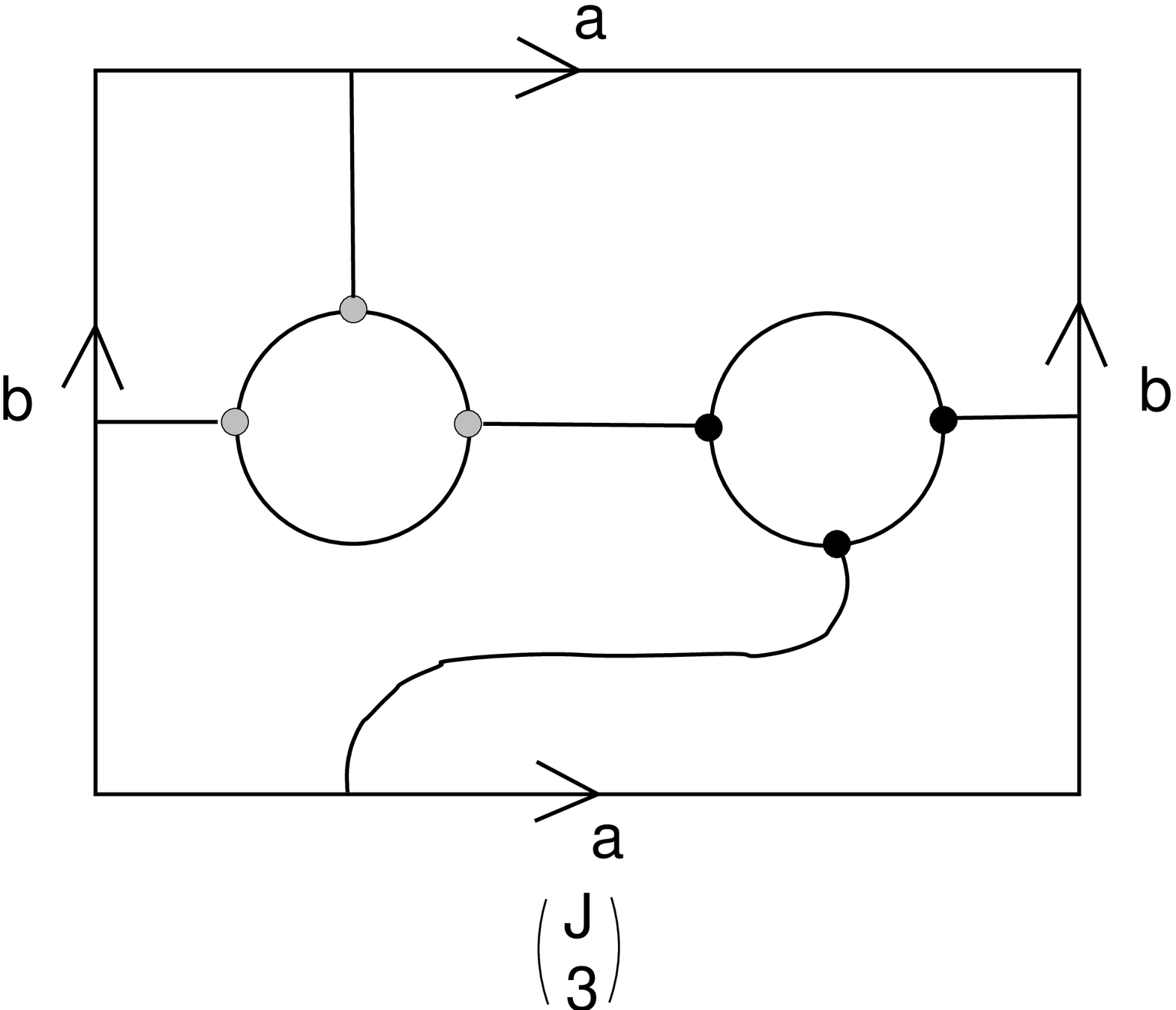}\qquad\qquad
\epsfxsize=4.5cm\epsfbox{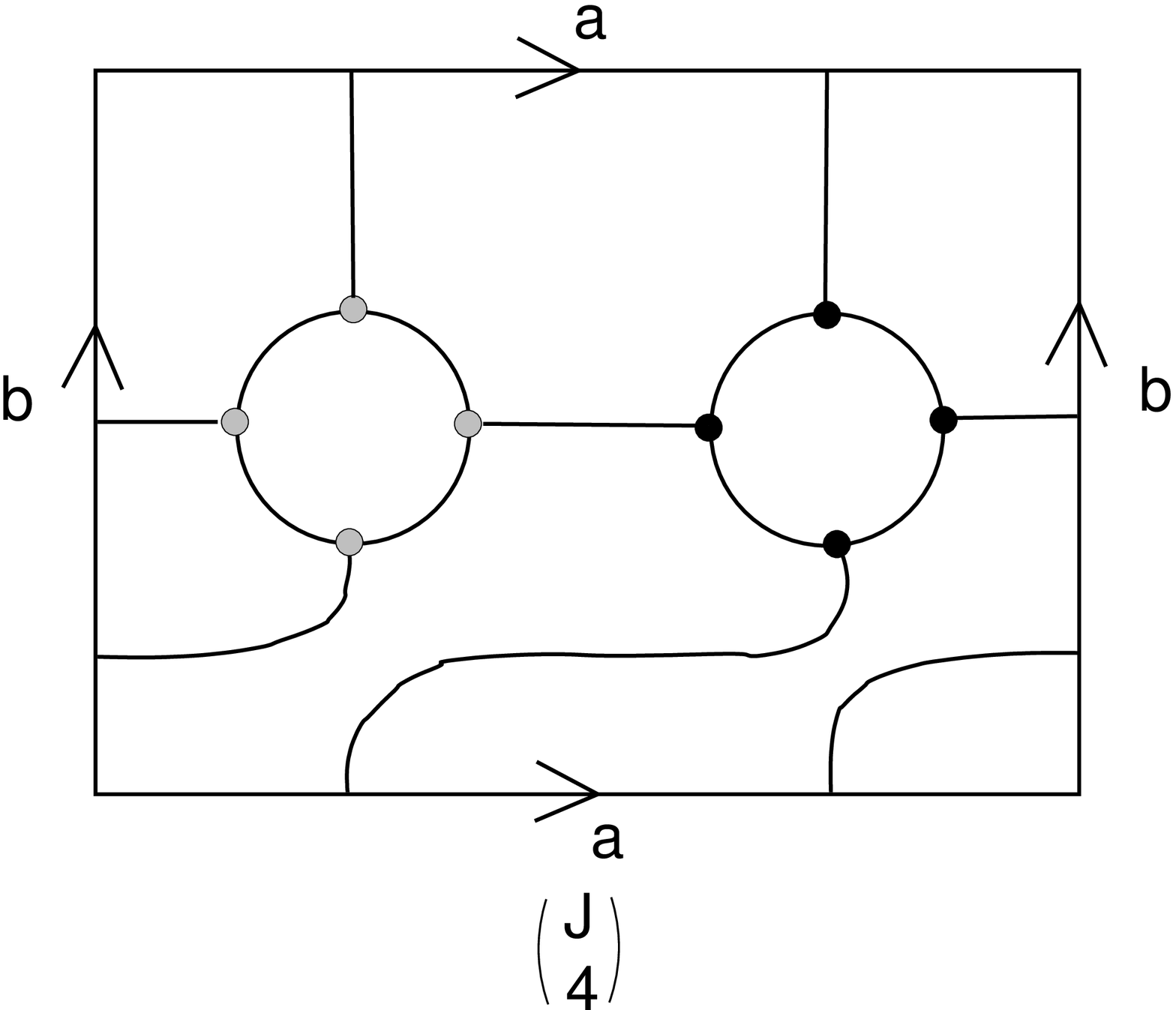}} }
\caption{}
\end{figure}

The structure of this result is easy do understand combinatorically.
We have to find the possible ways of connecting two necklaces with each,
respectively, $J$ white ($Z$'s) and $J$ black ($\bar Z$'s)
beads according to the following
rules: (a) each connection has to link a black to a white bead  
(b) in order to find the ${\cal O}(N^{J-2 h})$ contribution the 
connections have to be drawn without crossing on a genus $h$ 
surface such that no
handle of the surface can be collapsed without pinching a connection.
Let us call all connections that run (possibly after topological
deformation) parallel to another connection ``reducible''. Eliminate all
reducible connections. This will lead to a number of inequivalent,
irreducible graphs on the genus $h$ surface. These have $k$ connections
ranging between at least $k=2 h+1$ connections (easy to see) and
at most $k=4 h$ connections (harder to see). Even for a given number
of irreducible connections there are (starting from genus $h=2$)
an increasing number $a_{h,k}$ of inequivalent irreducible graphs. 
Once these numbers are
worked out (this is a formidable problem, except for genus 1, 
where $k=3,4$ and there is only one irreducible graph in each case, see
figure 1) the total number of graphs of irreduciblity type $k$ 
with $J$ connections is given by $a_{h,k} {J \choose k}$.
This explains the above expansion eq.(\ref{expansion}).
We are now ready to investigate the large $J$ limit of the correlator;
we see from eq.(\ref{expansion})
\begin{equation}
\langle {\rm Tr}Z^J~ {\rm Tr}\bar Z^J \rangle  = 
J N^J \Bigg\{ 1+ \sum_{h=1}^{\infty} 
{a_{h,4 h} \over (4 h)!} { J^{4 h} \over N^{2 h}} + \ldots \Bigg\}
\label{scaledexpansion}
\end{equation}
that the terms involving $a_{h,4h}$ stay finite in the double scaling limit
$J,N \rightarrow \infty$ iff we keep ${J^2 \over N}$ finite while
all other terms in eq.(\ref{scaledexpansion}) vanish.
This means that non-planar diagrams {\it survive} to all orders in
$1/N^2$. In addition the above analysis allows for the interpretation
of the scaling limit eq.({\ref{limit}) already mentioned in
the introduction: Clearly, in light of eq.(\ref{expansion}), 
the omission of the subleading terms eliminates diagrams with ``skinny''
handles consisting of only a small number of string bits.
The meaning of the double scaling limit is therefore that it
combines a continuum limit (taking the size of lattice strings to
infinity) with a large $N$ limit (ensuring that only ``fat'' handles
are allowed). This is very similar to the matrix model
double scaling limits discovered in \cite{Brezin:rb}. One apparent
difference is that in these ``old'' double scaling limits 
there is an exponential growth of the number of graphs at fixed genus.
However, this exponential factor (which in our case would be a term
like $c^J$) is well known to be non-universal, reestablishing the
close similarity.

It is straightforward to explicitly find the scaling function 
from eq.(\ref{corr})
\begin{equation}
G^{JJ'}(x)= \frac{\delta^{JJ'}}{|x|^{2J}}
\left(\frac{g_{{\rm YM}}^2N}{8\pi^2}\right)^J \cdot J\cdot \frac{
\sinh\left( \frac{1}{2}\frac{J^2}{N}\right) }{ \frac{1}{2}
\frac{J^2}{N} }
\end{equation}
incidentally yielding the numbers $a_{h,4 h}=(4 h)!/((2 h+1)! 2^{2 h})$.

At this point, one might raise the objection that the correct gauge
group for the AdS/CFT correspondence is SU(N) rather than U(N) and the
sub-leading orders in the large $N$ expansion differ for these two
groups.  Indeed, the analog of (\ref{corr}) can be found for SU(N),
\begin{eqnarray}
\lefteqn{ \langle {\rm Tr}Z^J~ {\rm Tr}\bar Z^J \rangle_{\rm SU(N)}
=}\nonumber \\ && J!N^{-J+2}(-1)^J+ \nonumber \sum_{p=1}^{J}\left(
\matrix{ J\cr p\cr} \right)^2 \frac{(-1)^{J-p} (J-p)! N^{p-J}} {(p+1)}\left(
\frac{\Gamma(N+p+1)}{\Gamma(N)} - \frac{\Gamma(N+1)}{\Gamma(N-p)}
\right)
\label{corr1}
\end{eqnarray}

It can be seen that this formula has the same asymptotics in large $J$
as the previous case of U(N), though it has slower rate of convergence
to the asymptote.  {}From this, we conclude that the large $J$ limit
of the U(N) and SU(N) gauge theories are similar enough that we can
focus on U(N).  Of course, this is true only for the scaling limit
that we are considering.  If, for example, $J$ goes to infinity faster
than $N^{1/2}$, the limit could be more complicated.

Other two-point functions easily evaluated (for $U(N)$) are
\begin{eqnarray}
\lefteqn{\langle 0| {\rm Tr}\left( Z^J(x)\Phi^3(x)\right) 
{\rm Tr}\left(\bar Z^{J'}(0)\Phi^3(0)\right) |0\rangle=}
\nonumber\\
&& \hspace{2.5cm}
\,\left(\frac{g_{{\rm YM}}^2 }{8\pi^2}\right)^{J+1}
\frac{\delta^{JJ'}}{|x|^{2J+2}}   
\left( \frac{\Gamma(N+J+2)}{\Gamma(N)} - \frac{\Gamma(N+1)}{\Gamma(N-J-1)}
 \right)
\nonumber\\
&& \hspace{2.5cm}\longrightarrow \, \frac{\delta^{JJ'}}{|x|^{2J+2}}
\left(\frac{g_{{\rm YM}}^2 N}{8\pi^2}\right)^{J+1} 
\cdot\frac{ \sinh\left( \frac{1}{2}\frac{J^2}{N} \right) }{
\frac{1}{2}\frac{J^2}{N} }
\end{eqnarray}
and
\begin{eqnarray} \lefteqn{\hspace{-.7cm}
\sum_{p=0}^J
\langle 0| {\rm Tr}\left( Z^{J-p}(x)\Phi^3(x) Z^p(x)\Phi^4(x)\right) 
{\rm Tr}\left(\bar Z^{J'-q}(0)\Phi^4(0)\bar Z^q(0)\Phi^3(0)
\right) |0\rangle= \nonumber}\\
\hspace{1.0cm}
&&\hspace{2.5cm}
 \frac{\delta^{JJ'}}{|x|^{2J+4}}\left(\frac{g_{{\rm YM}}^2}{8\pi^2}\right)^{J+2}
\left( \frac{\Gamma(N+J+2)}{\Gamma(N)} - \frac{\Gamma(N+1)}{\Gamma(N-J-1)} 
\right)
\nonumber\\
&&\hspace{2.5cm}
\longrightarrow \frac{\delta^{JJ'}}{|x|^{2J+4}}
\left(\frac{g_{{\rm YM}}^2 N}{8\pi^2}\right)^{J+2}
\cdot\frac{ \sinh\left( \frac{1}{2}\frac{J^2}{N} \right) }{
\frac{1}{2}\frac{J^2}{N} }
\end{eqnarray}
The last of these is a correlator of a chiral primary field with a
one which isn't a chiral primary.   This is due to the fact that the summation
effectively symmetrizes the operator product.  Since the sum depends on non-planar
diagrams, the terms individually must also.  This means that the quantity of
interest, $$\langle 0| {\rm Tr}\left( Z^{J-p}(x)\Phi^3(x) Z^{p}(x)\Phi^4(x)\right) 
{\rm Tr}\left(\bar Z^{q}(0)\Phi^4(0)\bar Z^{J-q}(0)\Phi^3(0)
\right) |0\rangle$$
must contain non-planar graphs in the limit (\ref{limit}).  
We shall discuss non-planar corrections to this correlator in 
section~\ref{momentum}. 

\sect{Three-point functions}

It is interesting to consider the three-point function of chiral
primary fields.  In the AdS/CFT correspondence, such three-point
functions should coincide with a 3-point amplitude for certain BPS
states of the graviton.  In the present case, we shall see that, like
the case of the two-point correlator of chiral primary operators, the
three-point function also obtains contributions from all genera in the
limit (\ref{limit}).  

Consider the three-point function of chiral primary operators
\bea
G^{JKL}(x,y)&=&\langle  0| {\rm Tr}\left(Z^J(x)\right) 
{\rm Tr}\left(Z^K(y)\right){\rm Tr}\left(\bar{Z}^L(0)\right)|0\rangle
\nonumber\\
&=& \delta^{L-J-K,0} \left(\frac{g^2_{{\rm YM}}}{8\pi^2}\right)^{J+K}
\frac{1}{|x|^{2J}|y|^{2K}}\,
\langle {\rm Tr} Z^J {\rm Tr} Z^K {\rm Tr} \bar{Z}^{J+K}\rangle
\eea
with \mbox{($0<J,K,L<N$)},
where again the remaining expectation value can be evaluated using 
matrix model techniques (cf.\ appendix A)
\bea
\langle {\rm Tr} Z^J {\rm Tr} Z^K {\rm Tr}\bar{Z}^{J+K}\rangle
&=&\frac{1}{J+K+1}
\left\{
\frac{\Gamma(N+J+K+1)}{\Gamma(N)}+\frac{\Gamma(N+1)}{\Gamma(N-J-K)}
\right.
\nonumber \\
&&
-\left.\frac{\Gamma(N+J+1)}{\Gamma(N-K)}-\frac{\Gamma(N+K+1)}{\Gamma(N-J)}
\right\} \label{tpoint}
\eea
The scaling limit of the three-point function is
\bea
G^{JKL}(x,y)&=&2\, \delta^{L-J-K,0}\,
(J+K)\left(\frac{g^2_{{\rm YM}}N}{8\pi^2}\right)^{J+K}
\frac{1}{|x|^{2J}|y|^{2K}}\, \cdot \\
&&\frac{\sinh\left(\frac{J(J+K)}{2N}\right)\sinh\left(\frac{K(J+K)}{2N}\right)}
{\frac{(J+K)^2}{2N}}
\eea

We see that all genera contribute.  Furthermore, the dependence of
these functions on the parameters $J^2/N,K^2/N,L^2/N$ is one which
cannot be removed by normalizing the operators.  This implies a
dependence of the three-point amplitude on the three
parameters which are left over in the double-scaling limit.

\sect{Computing momentum correlators \label{momentum}}

We consider the following operators
\be
\cO_n(x) = \frac{1}{\sqrt{J}\, N^{J/2+1}}\, \sum_{l=0}^J \Tr [ \, \phi_3
\, Z^l\, \phi_4\, Z^{J-l}\, ]\, e^{\frac{2\pi\, i \, n\, l}{J}}
\label{momentumop}
\ee 
Notice that these operators differ from those introduced 
in~\cite{Berenstein:2002jq} by the inclusion of $l=0$ in the summation
range. This difference will turn out to be important for what follows.
We shall be interested in the two point correlator $\langle 
 \cO_{n_1}(x)\,\bar\cO_{n_2}(0) \rangle$ up to next to leading
order in $N^2$. 
In the free theory limit
one has after contracting the $\phi_3$ and $\phi_4$'s
\bea
\lefteqn{
\langle \,\cO_{n_1}(x)\,\bar\cO_{n_2}(0)\, \rangle_{0-loop} =}\nonumber \\ 
&&
\frac{1}{J\, N^{J+2}}\,\left(\frac{g_{{\rm YM}}^2}{8\pi^2|x|^2}\right)^{J+2} 
\sum_{p,q=0}^{J}\,\langle \, \Tr [ \, Z^{J-p}\, \Zb^{J-q}\, ]\,
\Tr [ \, Z^{p}\, \Zb^{q}\, ]\,\rangle\,
 e^{\frac{2\pi\, i}{J} (n_1\, p-n_2\, q)}
\label{momcorr}
\eea
The remaining correlator splits into a disconnected and
connected piece
\be
\langle \, \Tr [ \, Z^{J-p}\, \Zb^{J-q}\, ]\,
\Tr [ \, Z^{p}\, \Zb^{q}\, ]\,\rangle =
\delta_{p,q}\, [J\mid p] + [J\mid p,q\,]
\label{FullJpq}
\ee
where we have defined 
\bea
[J\mid p\,]&=&\langle  \Tr [ \, Z^{J-p}\, \Zb^{J-p}\, ]\,\rangle\, 
\langle  \Tr [ \, Z^{p}\, \Zb^{p}\, ]\,\rangle\nn\\
\relax
[J\mid p,q\,] &=& \langle 
\Tr [ \, Z^{J-p}\, \Zb^{J-q}\, ]\,
\Tr [ \, Z^{p}\, \Zb^{q}\, ]\,\rangle_{conn} 
\eea
The disconnected piece may easily be deduced from \eqn{corr} by making use of
the
identity
\be
\langle{\rm Tr}(Z^{J+1}) {\rm Tr}(\bar{Z}^{J+1})  \rangle
=(J+1)\langle {\rm Tr} (Z^J \bar{Z}^J)\rangle
\ee
The result reads
\be
[J\mid p]
=
N^{J+2} + N^{J}\, \Bigr [ \binomial{J-p+2}{4}+\binomial{p+2}{4}\, \Bigl ]
\, +\, \cO(N^{J-2})
\label{Jp}
\ee
The combinatorics of the connected contributions goes as follows.
As the leading disconnected contribution to \eqn{FullJpq} scales as
$N^{J+2}$ we shall be interested only in the planar connected contribution
to $[J\mid p,q\,]$. The relevant contractions are depicted in 
figure~\ref{fig:contjpq}.
Assume $J-q>p$ and $q>p$. Then we connect $k$ of the $(J-q)$
$\Zb$'s with $k$ of the $p$ $Z$'s. There are $(p-k+1)\, (J-q-k+1)$
ways of doing this and we should sum over these $k$ contractions
as $k=1,\ldots, p$. All the remaining contractions (depicted by dashed
lines in the figure) are completely determined by planarity once the middle
$k$ lines have been chosen. If we have no middle contractions
($k=0$) all the $p$ $Z$'s are contracted with $p$ $\Zb$'s on the same 
ellipse, leaving
$(q-p)$ $\Zb$'s on the right ellipse of figure~\ref{fig:contjpq} 
to be contracted
with the left ellipse. There are then $(q-p)\, (p+1)\, (J-q+1)$
ways of doing this. We hence see that
\bea
[J\mid p,q\,] &=&
N^J\, \Bigl [ \, \frac 1 6\, p\, (p+1)\,(3J+1-p-3q) \nn\\
&& \qquad +
(q-p)\, (p+1)\, (J-q+1)\, \Bigr ] \, +\, \cO(N^{J-2})\nn\\
&&\mbox{for}\quad (q>p\, , \,J-q>p)
\label{Jpq}
\eea

\begin{figure}[t]
\centerline{\epsfxsize=7cm\epsfbox{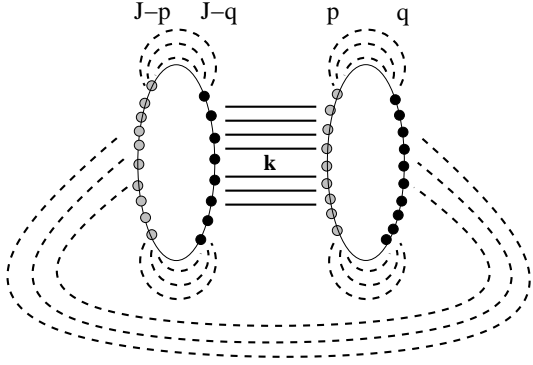}} 
\caption{The contractions \label{fig:contjpq}  }
\end{figure}

For different regions of $p$ and $q$ the correlator is determined
by the two obvious symmetries $(p\leftrightarrow q)$ and 
$(p\rightarrow J-p, q\rightarrow J-q)$ of $[J\mid p,q\, ]$. One may
now Fourier transform \eqn{Jp} and \eqn{Jpq} and deduce the leading
behaviours in $J^2/N$. For the disconnected contribution one finds
\bea
\lefteqn{D_{n_1-n_2}=
\sum_{p=0}^J\, [J\mid p\, ]\, e^{\frac{2\pi\, i}{J} (n_1-n_2)\, p}=}\nn\\
\lefteqn{\qquad N^{J+2}\, (J+1)\, \delta_{n_1,n_2}+}\nn\\
&&
\hspace{-0.5cm}
 \qquad + N^J\, J^5\, 
\cases{\frac{1}{60} & $n_1=n_2$\cr 
\frac{8 \pi^2 n_1^2-3}{384 n_1^4 \pi^4}& $n_1=-n_2$ \cr
\frac{2\pi^2(n_1-n_2)^2-3}{24(n_1-n_2)^4\pi^4}& $n_1\neq n_2$\cr}
\quad + N^J\, \cO(J^4) + \cO(N^{J-2})
\eea
in the limit of large $J$.
Let us now turn to the connected non-diagonal
contribution to the correlator \eqn{momcorr}.
The summation domain for $p$ and $q$ then splits into
four sectors $(q>p,J-q>p),(q>p,J-q<p),(q<p,J-q>p)$ and
$(q<p,J-q<p)$ and due to the symmetries of $[J\mid p,q\,]$ the
sum over these four sectors may be reduced to a single sum over
the domain $(q>p,J-q>p)$ with varying phase factor 
contributions. Explicitly one has to evaluate the double sum
\be
C_{n_1,n_2}=
2\,\sum_{p=0}^{[J/2]-1}\sum_{q=p+1}^{J-p-1}[J\mid p,q\, ]\,
\Bigl ( \cos[\ft{2\pi}{J}(p\, n_1-q\, n_2)]+
\cos[\ft{2\pi}{J}(q\, n_1-p\, n_2)]\, \Bigr )
\label{doublesum}
\ee
where $[J/2]=J/2$ for $J$ even and $[J/2]=(J+1)/2$ for J odd. 
As $[J\mid p,q\,]$ scales as $p^3$ the double sum \eqn{doublesum}
will scale as $J^5$ and its contribution is relevant to the momentum
correlator in the double scaling limit. 
Performing the sums in \eqn{doublesum} and taking the
$J\rightarrow \infty$ limit we find 
\be
C_{n_1,n_2}= N^J\, J^5\, 
\cases{\frac{1}{40} & $n_1=n_2=0$\cr
\frac{3-2\pi^2\, n^2}{24\, n^4\pi^4}& $n:=n_{1/2}\neq0,n_{2/1}= 0$ \cr
\frac{21-2\pi^2\,n^2}{48\,n^4\pi^4}& $n:=n_1=n_2$ and $n_1\neq0\neq n_2$
\cr
\frac{9}{32 n^4 \pi^4}&  $n:=n_1=-n_2$ and $n_1\neq0\neq n_2$ \cr
\frac{2\, n_1{}^2-3\,n_1\,n_2+2\, n_2{}^2}
{8\,n_1{}^2\, n_2{}^2\,(n_1-n_2)^2\,\pi^4}& $n_1\neq n_2$ and 
$n_1\neq0\neq n_2$ \cr}
\ee
up to terms of order $N^J\, \cO(J^4)$ and $\cO(N^{J-2})$.
We now have to worry about the contributions along the ``diagonals''
$p=q$ and $J-q=p$  
which were omitted in the double sum of 
\eqn{doublesum}. The counting now works analogously, for the 
first case ($p=q$) one has
\be
C^{p=q}_{n1-n2}=N^J\,\sum_{p=0}^{[J/2]}\ft 1 6\, (p+1)\, p\, (3J+1-4p)\,
2\, \cos[\, \ft{2\pi}{J}\, (n_1-n_2)p\, ]
\ee
and for the second case ($J-q=p$) one finds
\be
C^{J-q=p}_{n1-n2}=N^J\,\sum_{p=0}^{[J/2]}\, \Bigl[
\ft 1 6\, (p+1)\, p\, (2p+1) + (p+1)^2\, (J-2p)\, \Bigr ]\,
2\, \cos[\, \ft{2\pi}{J}\, (n_1-n_2)p\, ]
\ee

These sums can be worked out explicitly, however, we already see at this
point that they will not contribute at order $J^5$ as the summand
scales as $p^3$.
We hence have that
\be
C^{p=q}_{n1-n2} \sim C^{J-q=p}_{n1-n2} \sim N^J\, \cO(J^4)
\ee
and these contributions are suppressed in the limit \eqn{limit}.
Summarizing we have thus found that
\be
\label{0loop}
\langle \,\cO_{n_1}(x)\,\bar\cO_{n_2}(0)\, \rangle_{0-loop} = 
\left(\frac{g_{{\rm YM}}^2 }{8\pi^2|x|^2}\right)^{J+2}\, \Bigr [ \, \delta_{n_1
n_2}+
\frac{J^4}{N^2}\, M_{n_1 n_2} + \cO(\frac{J^3}{N^2})
\, \Bigr ]
\ee
where we have introduced the symmetric real matrix $M_{\bar n \bar m}$ given by
\be
M_{00}=\frac{1}{24}\qquad M_{0\bar n}=0\qquad
M_{\bar n \bar m}= 
\cases{ \frac{2\pi^2(\bar n-\bar m)^2-3}{24(\bar n-\bar m)^4\pi^4} +
\frac{2\, \bar n^2-3\,\bar n\,\bar m+2\, \bar m^2}
{8\,\bar n^2\, \bar m^2\,(\bar n-\bar m)^2\,\pi^4}
& $|\bar n |\neq |\bar m|$\cr
& \cr
\frac{1}{60} + \frac{21-2\pi^2\,\bar n^2}{48\,\bar n^4\pi^4}
& $\bar n = \bar m$\cr
& \cr
\frac{105+8 \pi^2 \bar n^2}{384\, \bar n^4\, \pi^4}
& $\bar n = -\bar m$\cr}
\ee
Note the decoupling of the zero momentum sector
($M_{0\bar n}=0$) which corresponds to the groundstate not mixing 
under perturbation theory  with the excited states in the dual string
picture.
This result tells us that we have to renormalize our momentum
operators according to
\be
\tilde\cO_n(x)= \sum_m \Bigr [ \, \delta_{n m}-
\frac{J^4}{2\, N^2}\, M_{n m} \, \Bigr ]\, \cO_m(x)
\label{redefinedops}
\ee
in order to maintain orthogonality up to order $J^4/N^2$, i.e.
$\langle  \,\tilde \cO_n\, \tilde \cO_m\,\rangle \propto
\delta_{n,m}$.

\sect{Computing anomalous dimensions}

In this section we shall review the computation of anomalous
dimensions of operators which are of interest to us in this paper.
The correlation functions of interest are those of traces of products
of the scalar fields, for example
\bea
\lefteqn{
\langle 0| :{\rm Tr}\Phi_{\alpha_1}(x)\ldots \Phi_{\alpha_p}(x):
:{\rm Tr}\Phi_{\beta_1}(0)\ldots \Phi_{\beta_q}(0): |0\rangle =}\nonumber
\\
&& \hspace{0.5cm}
\delta_{pq}g_{{\rm YM}}^{2p}\Delta^p(x)\sum_{\rm Perm~P}
\left(\delta_{\alpha_1\beta_{P(1)}}\ldots\delta_{\alpha_p\beta_{P(p)}}\right)
+{\rm quantum~corrections}
\eea
Here, the dots $:...:$ indicate normal ordering, so that the vacuum
expectation value of each individual operator vanishes. In the
following, we will omit the normal ordering symbols.  The first term
on the right-hand-side is the correlator in the free field limit.  The
other terms, are given by radiative corrections to the free field
limit.  In this section, we will compute these radiative corrections
to the next order in $g_{{\rm YM}}^2$.

\subsection{Notation \label{notation} }

The field content of ${\cal N}=4$ supersymmetric Yang-Mills theory in
four dimensions are the scalars, $\Phi_\alpha(x)$ where the Greek
indices $\alpha,\beta,\ldots=1,...,6$ transform under the R-symmetry
SO(6)$\sim$SU(4), vectors $A_\mu(x)$ where $\mu=1,2,3,4$ is the
space-time index and a sixteen component spinor $\Psi(x)$.  These
fields are Hermitean $N\times N$ matrices and for the most part in the
following, the gauge group will be U(N).  These fields can be expanded
in terms of the generators $T^a$ of U(N) as
\begin{equation}
\Phi_\alpha(x)=\sum_{a=0}^{N^2-1} \phi^a_\alpha(x)T^a ~~,~~
A_\mu(x)=\sum_{a=0}^{N^2-1} A_\mu^a(x)T^a  ~~,~~
\Psi(x)=\sum_{a=0}^{N^2-1} \psi^a(x)T^a
\end{equation}
where $T^0_{kl}=\frac{1}{\sqrt{2N}}\delta_{kl}$ is the U(1) generator.
We use the letters \mbox{$a,b,\ldots=0,\ldots ,N^2-1$} to denote components
in the Lie algebra of U(N).
The conventions for the generators and structure constants are
$$
\left[ T^a, T^b\right]=if^{abc}T^c
~,~~
{\rm Tr} (T^a T^b)=\frac{1}{2}\delta^{ab}
$$
Also, 
$$
\sum_{a=0}^{N^2-1} T^a_{ij}T^a_{kl}= \frac{1}{2}\delta_{il}\delta_{jk}
~,~~
\sum_{a=1}^{N^2-1}  T^a_{ij}T^a_{kl}=
\frac{1}{2}\left(\delta_{il}\delta_{jk}-
\frac{1}{N}\delta_{ij}\delta_{kl}\right)
$$
and
$$
f^{abc}f^{a'bc}=N\delta^{aa'}
~,~~
f^{abc}f^{abc}=N(N^2-1)
$$
We shall also need the identity
\bea
\lefteqn{\Tr(T^a\, A\, T^b\, B)\,\Tr(T^c\, C\, T^d\, D)\, f^{ead}\,
f^{ebc} =}
\label{identity1}\\
&&\ft 1 8 \Bigl ( \Tr\, A\, \Tr\, C\, \Tr\, BD +\Tr \, B\, \Tr\, D\, \Tr\,
AC
-\Tr\, ACBD\,- \Tr \, ADBC\, \Bigr )\nn
\eea
along with
\bea
\Tr (T^{\bar a}\, A)\, \Tr (T^{\bar a}\, B) &=& \ft 1 2\, \Tr \, AB -\ft
{1}{2N}\,
\Tr\, A\, \Tr\, B \nn\\
\Tr (T^{\bar a}\, A\, T^{\bar a}\, B) &=& 
\ft 1 2\, \Tr \, A\, \Tr\, B -\ft {1}{2N}\,
\Tr\, AB 
\label{identity2}
\eea
where $A,B,C,D$ are $N\times N$ matrices and
$\bar a=1,\ldots,N^2-1$ runs only over the SU(N) indices. We use 
the conventions of \cite{Erickson:2000af,Plefka:2001bu}.

The Euclidean space action of ${\cal N}=4$ supersymmetric Yang-Mills
theory is
\begin{eqnarray}
S=\int d^4x \frac{1}{g^2_{{\rm YM}}}\left\{
\frac{1}{4}F_{\mu\nu}^aF_{\mu\nu}^a
+\frac{1}{2}(D_\mu\phi_\alpha^a)^2+\frac{1}{4}\sum_{\alpha,\beta=1}^{6}
f^{abc}f^{ade}\phi^b_\alpha\phi^c_\beta\phi^d_\alpha\phi^e_\beta +
\right.  \nonumber \\ \left.  +\frac{1}{2}\bar\psi^a\Gamma^\mu
D_\mu\psi^a+\frac{1}{2}f^{abc}\bar\psi^a\Gamma^\alpha
\phi^b_\alpha\psi^c \right\}
\label{symaction}
\end{eqnarray}
where the curvature is $F_{\mu\nu}^a=\partial_\mu A_\nu^a-\partial_\nu
A_\mu^a + f^{abc}A_\mu^b A_\nu^c$ and the covariant derivative is
$D_\mu\phi^a= \partial_\mu\phi^a+ f^{abc}A^b_\mu\phi^c$.
$(\Gamma^\mu,\Gamma^\alpha)$ are the ten-dimensional Dirac matrices in
the Majorana-Weyl representation.

Vertices and propagators can be read off from the action
(\ref{symaction}).  We will work in the Feynman gauge where the free field 
limit of the vector field propagator is
$$
\langle 0| A^a_\mu(x) A_\nu^b(0) |0\rangle_0=
g_{{\rm YM}}^2\delta^{ab}\delta_{\mu\nu}\frac{1}
{4\pi^2[x]^2}
$$
In this gauge, it resembles the free scalar field propagator
$$
\langle 0| \phi^a_\alpha(x)\phi^b_\beta(0) |0\rangle_0=
g_{{\rm YM}}^2 \delta^{ab}\delta_{\alpha\beta}\frac{1}{4\pi^2[x]^2}
$$

${\cal N}=4$ supersymmetric Yang-Mills theory contains no dimensional
parameters.  Furthermore, because of the high degree of supersymmetry,
it has vanishing beta function for the coupling constant $g_{{\rm YM}}$ and
is a conformal field theory.  However, individual Feynman diagrams are
divergent and the explicit computations that we do in the following
will require a regularization.  For this purpose, we will use
regularization by dimensional reduction.  Four dimensional ${\cal
N}=4$ supersymmetric Yang-Mills theory is a dimensional reduction of
the ten-dimensional ${\cal N}=1$ theory.  We can obtain a
regularization of the four dimensional theory which still has sixteen
supersymmetries (but of course is no longer conformally invariant) by
considering a dimensional reduction to $2\omega$ dimensions, rather
than 4 dimensions.  In this reduction, the fermion content is still a
sixteen-component spinor (which originally was the Weyl-Majorana
spinor in ten dimensions).  There are $10-2\omega$ rather than $6$
flavors of scalar field and $2\omega$ rather than $4$ polarizations of
the vector field.

The Green function for the Laplacian in $2\omega$ dimensions is
\begin{equation}
\Delta(x)= \int \frac{ d^{2\omega} p} {(2\pi)^{2\omega} } \frac{e^{ip\cdot
x}}
{ p^2 } = \frac{ \Gamma(\omega-1) }{4\pi^\omega [x^2]^{\omega-1} }
\end{equation}
When we work in $2\omega$-dimensions, the right-hand side of this
equation should replace $1/4\pi^2[x^2]$ in propagators.
Useful formulae for Feynman integrals in $2\omega$-dimensions can be
found in refs.\cite{Erickson:2000af,Plefka:2001bu}.

\subsection{One-loop integrals}

The Feynman diagrams which contribute to radiative corrections to
two-point correlators of scalar trace operators are depicted in
figure 3.  The first diagram in figure 3 contains the self-energy of
the scalar field.  Feynman rules and conventions for vertices can be
deduced from the action (\ref{symaction}).

The self-energy of the scalar field was computed with the present
notational conventions in refs.\cite{Erickson:2000af,Plefka:2001bu}.
Written as a correction to the scalar propagator, it is
\bea
\lefteqn{
\langle 0| \phi^a_\alpha(x) \phi^b_\beta(0) |0\rangle 
= }\nonumber \\
&& \hspace{0.5cm}g_{{\rm YM}}^2
\delta_{\alpha\beta} \Delta(x)\left( \delta^{ab}
+\delta^{\bar a\bar b}
\left[ -\frac{ g_{{\rm YM}}^2N \Gamma(\omega-1) }{
8\pi^\omega (2-\omega) (2\omega-3) }\right]
[x^2]^{(2-\omega)}+\ldots\right)
\label{selfenergy}
\eea
Note that the quantum correction only affects the SU(N) indices.  The
$\ldots$ denote terms of order at least $g_{{\rm YM}}^6$.

\begin{figure}[t]
\begin{center}
\begin{tabular}{ccccc}
\raisebox{.3cm}{{\epsfxsize=3.5cm\epsfbox{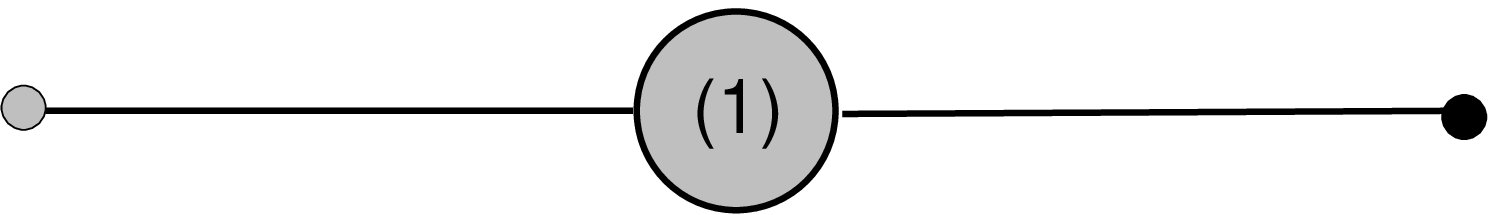}}} & &
\raisebox{-.15cm}{{\epsfxsize=3.5cm\epsfbox{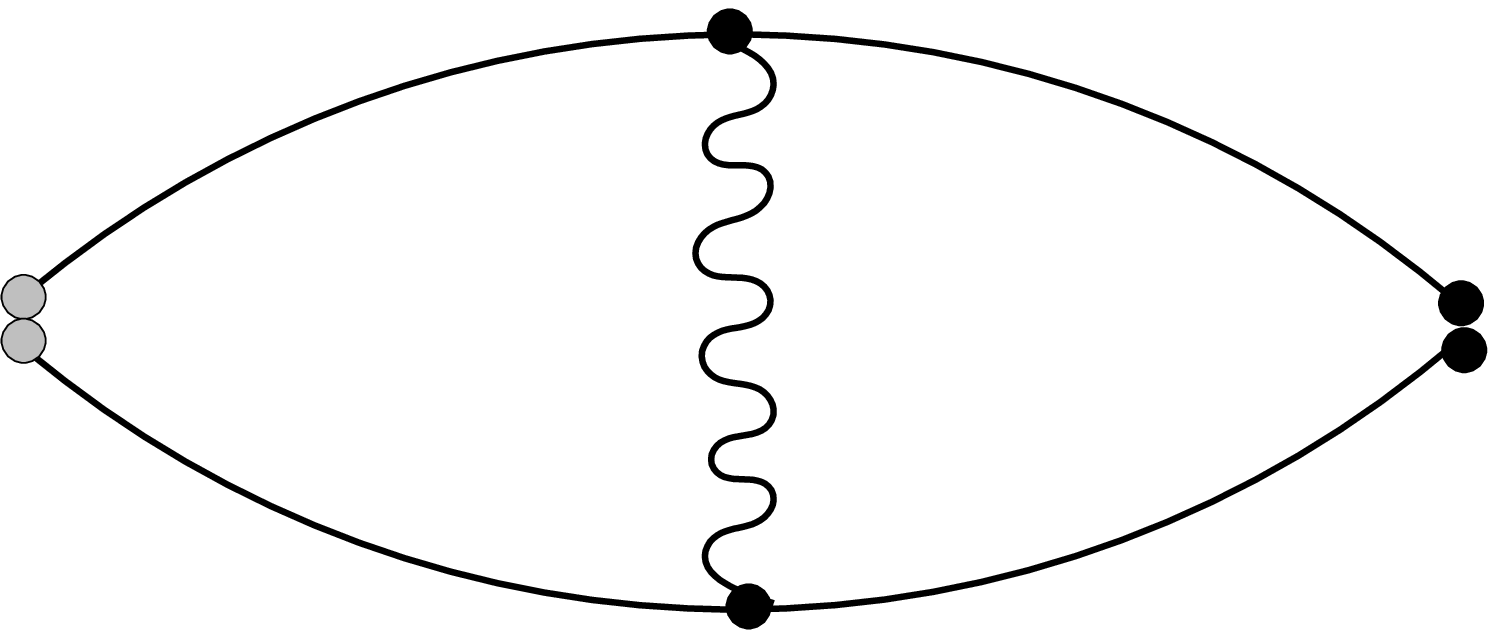}}} & &
{\epsfxsize=3.5cm\epsfbox{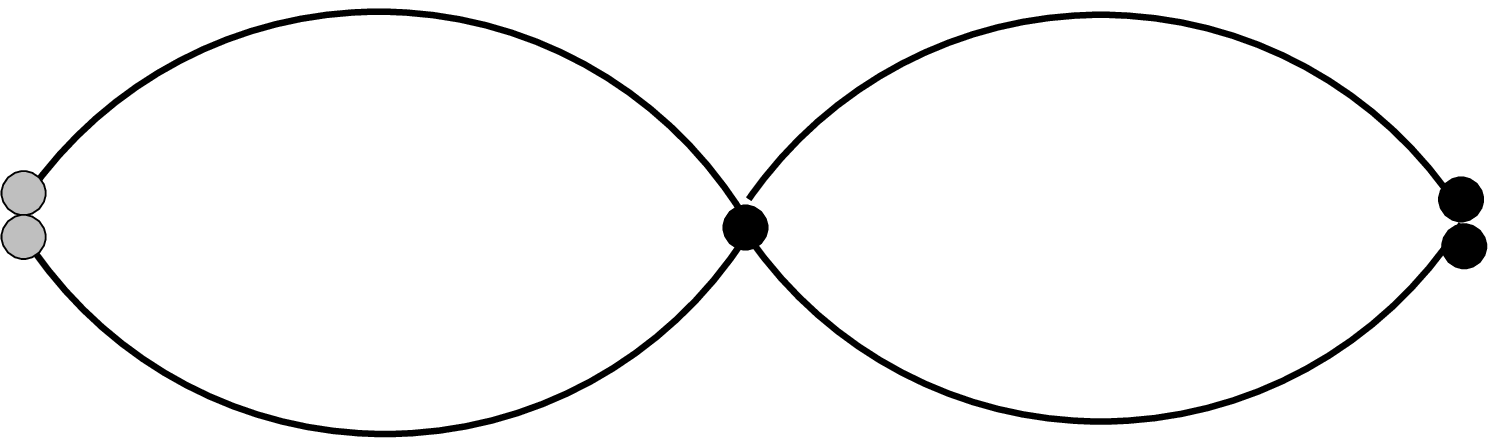}} \cr
\end{tabular}
\caption{The relevant graphs of the self energy, gluon exchange and 
four-point interaction contributing to the leading radiative corrections
of two-point scalar field trace operators.}
\end{center}
\end{figure}

The other quantum correction that we must take into account occurs in
the correlator of four scalar fields,
\begin{eqnarray}
\Gamma^{abcd}_{\alpha\beta\gamma\delta}(x)&\equiv &
\langle 0|
\phi^a_\alpha(x) \phi^b_\beta(x) \phi^c_\gamma(0)\phi^d_\delta(0)
|0\rangle  
%~~~~~~~~~~~~~~~~~~~~~~~~~~~~~~~~~~~~~~~~~~ 
\nonumber \\
&=& g_{{\rm YM}}^4\Delta^2(x)~
\left( \delta^{ac}\delta^{bd}\delta_{\alpha\gamma}\delta_{\beta\delta}
+\delta^{ad}\delta^{bc}\delta_{\alpha\delta}\delta_{\beta\gamma}\right)
+ \delta \Gamma^{abcd}_{\alpha\beta\gamma\delta}(x)
\label{g}
\end{eqnarray}
The first term on the right-hand-side is the free field limit and
$\delta \Gamma^{abcd}_{\alpha\beta\gamma\delta}(x)$ 
denotes the radiative corrections, which we shall compute to
order $g_{{\rm YM}}^6$. The relevant Feynman diagrams are the second and
third diagrams in figure 3.  When these diagrams are combined, the
total contribution is conveniently summarized in a combination of
three different tensor structures,
\begin{eqnarray}
\lefteqn{\delta \Gamma^{abcd}_{\alpha\beta\gamma\delta}(x)=
g^6_{\rm YM} \times} \nn\\&
\Big[ \left( \delta_{\alpha\gamma}\delta_{\beta\delta}f^{\small eac}f^{\small
ebd} 
+ \delta_{\alpha\delta}\delta_{\beta\gamma}f^{\small ead}f^{\small
ebc}\right)
\left[ \frac{ \Gamma(\omega-1)}{ 4\pi^\omega (2-\omega)(2\omega-3) }
\right] [x^2]^{2-\omega} \Delta^2(x) +
\label{symmetric} \\&
\left( \delta_{\alpha\delta}\delta_{\beta\gamma}-\delta_{\alpha\gamma}
\delta_{\beta\delta} \right) f^{\small eab}f^{\small ecd}
\left[ \frac{ \Gamma^2(\omega-2)\Gamma^2(\omega-1)\Gamma(3\omega-4) }
{ 16\pi^\omega\Gamma^2(2\omega-2)\Gamma(4-2\omega) } \right] 
[x^2]^{2-\omega}
\Delta^2(x) +
 \label{antisymmetric}\\&
\delta_{\alpha\beta}\delta_{\gamma\delta}\left(f^{\small eda}f^{\small
ecb}+
f^{\small edb}f^{\small eca}\right) 
\left[ \frac{ \Gamma^2(\omega-2)\Gamma^2(\omega-1)\Gamma(3\omega-4) }
{ 16\pi^\omega\Gamma^2(2\omega-2)\Gamma(4-2\omega) } \right] 
[x^2]^{2-\omega} 
\Delta^2(x) \Big]
\label{diagonal}
\end{eqnarray}
Here, we have dropped a possible contact term which, in four
dimensions, is proportional to the Dirac delta function $\delta(x)$
and therefore cannot contribute.
The terms are arranged so that the first part
(\ref{symmetric}) is symmetric, the second term (\ref{antisymmetric})
is anti-symmetric and the third term (\ref{diagonal}) is diagonal in
the indices $\alpha$ and $\beta$.  We caution the reader that the
symmetric term (\ref{symmetric}) is not traceless.  However, it is the
only one which will be needed when we compute the leading radiative
corrections to a correlator of chiral primary operators.

\subsection{Cancellation of leading radiative corrections to 2-point 
correlator of chiral primary operators}

The cancellation of leading order corrections to two- and three-point
functions of chiral primary operators was demonstrated in
ref.\cite{D'Hoker:1998tz}.  Here, to check our own procedure, and
further develop the matrix model approach to computations we will redo
this check of the no-renormalization theorem.

We will consider the two-point correlator of chiral primary operators,
\begin{equation}
\langle  0| {\rm Tr} Z^J(x) ~{\rm Tr} \bar{Z}^J(0) |0\rangle 
=\frac{\Delta^J(x)}{2^J}\cdot 
\langle {\rm Tr} Z^J ~{\rm Tr} \bar{Z}^J\rangle+\ldots
\end{equation}
The first term on the right-hand-side is the free field limit which is
given by the Feynman diagram depicted in figure \ref{fig:free}.  The
last bracket in this term is the matrix model correlation function in
eqn.(\ref{mmodel}).  The factor $\Delta^J(x)$ is the $J'$th power of
the scalar correlation function and the combinatorics of taking traces
of Lie algebra generators in the appropriate permutations is solved by
the matrix model correlator, which we have evaluated explicitly in
eqn.(\ref{corr}).

\begin{figure}[htb]
\centerline{\epsfysize=2cm\epsffile{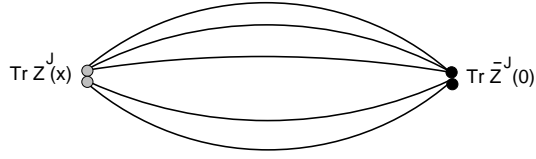}}
\caption{Feynman diagram for the free field limit of the correlator 
$\langle 0|{\rm Tr}Z^J(x)\, {\rm Tr} \bar{Z}^J(0) |0\rangle $.
There are $J$ scalar propagators connecting the points $x$ and $0$. 
\label{fig:free}}
\end{figure}

Radiative corrections to this result arise from the Feynman diagrams,
typical ones of which are depicted in figure \ref{fig:radcorr}.  The
first of the diagrams represent the radiative correction coming from
the self-energy of the scalar field.  As we discussed in the previous
subsection, this gives a contribution consisting of a factor of the
scalar self-energy times the number of scalar lines in the diagram.
Thus, the correction from self-energies is
$$
\left[ -\frac{ g^2_{{\rm YM}}N \Gamma(\omega-1) }{ 8\pi^\omega (2-\omega)
(2\omega-3) }\right] \cdot[x^2]^{(2-\omega)} 2^{-J} \Delta^J(x)\cdot
J^2\langle {\rm Tr}T^{\bar a} Z^{J-1} ~ {\rm Tr}T^{\bar a} Z^{J-1}
\rangle
$$
where we have taken into account that the interaction affects only the
SU(N) part of the propagator by inserting the SU(N) generators
$T^{\bar a}$ into the traces and eliminating one factor of $Z$ and
$\bar Z$ from each trace, respectively.  The factor of $J^2$ reflects
the fact that there are $J$ positions at which $T^{\bar a}$ could be
inserted in each product.  This equation can be simplified using
(\ref{identity2}) to get
\begin{equation}
\left[ -\frac{ g^2_{{\rm YM}}N \Gamma(\omega-1) }{ 8\pi^\omega (2-\omega)
(2\omega-3) }\right] \cdot[x^2]^{(2-\omega)} 2^{-J}
\Delta^J(x)\frac{J^2}{2}\langle {\rm Tr}Z^{J-1}\bar Z^{J-1}-
\frac{1}{N}{\rm Tr}Z^{J-1}{\rm Tr}\bar Z^{J-1} \rangle
\label{self-energy}
\end{equation}

\begin{figure}[htb]
\centerline{ \epsfysize=1.7cm\epsffile{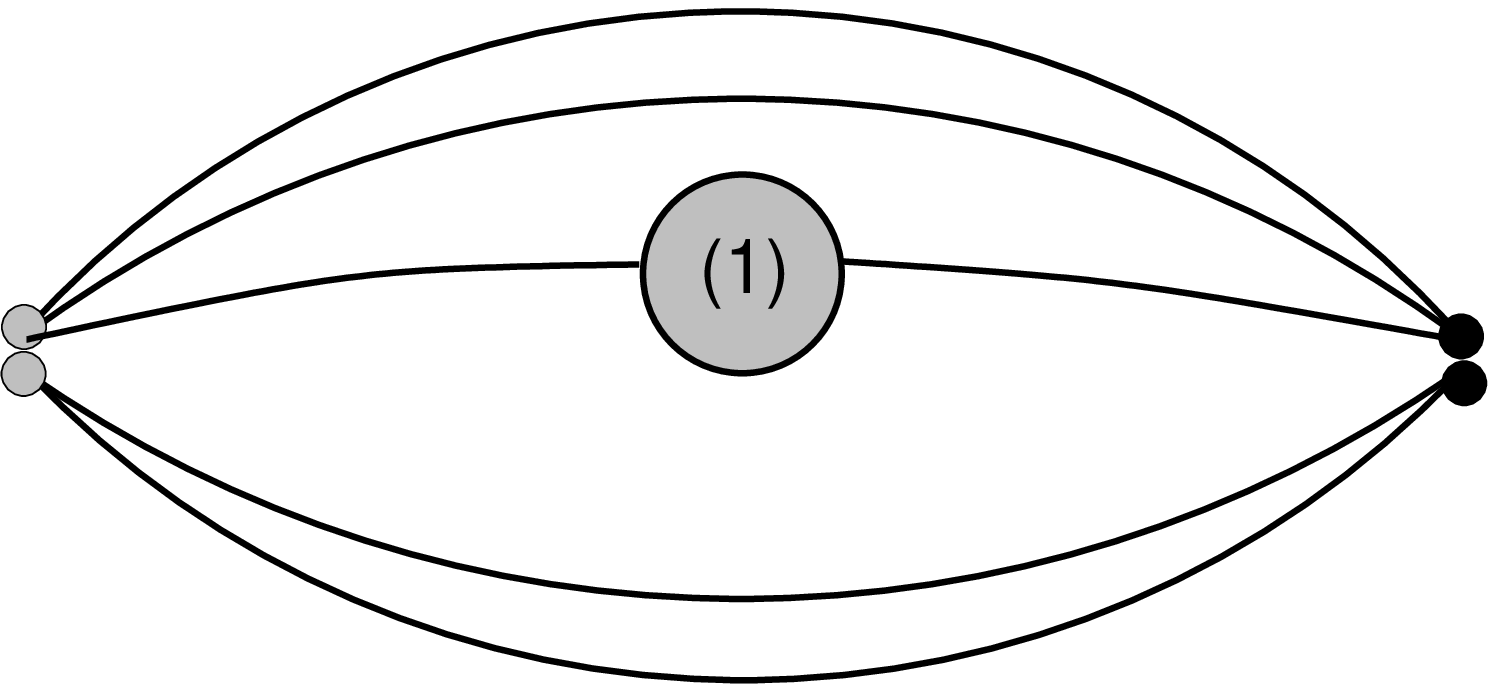}\quad
\epsfysize=1.7cm\epsffile{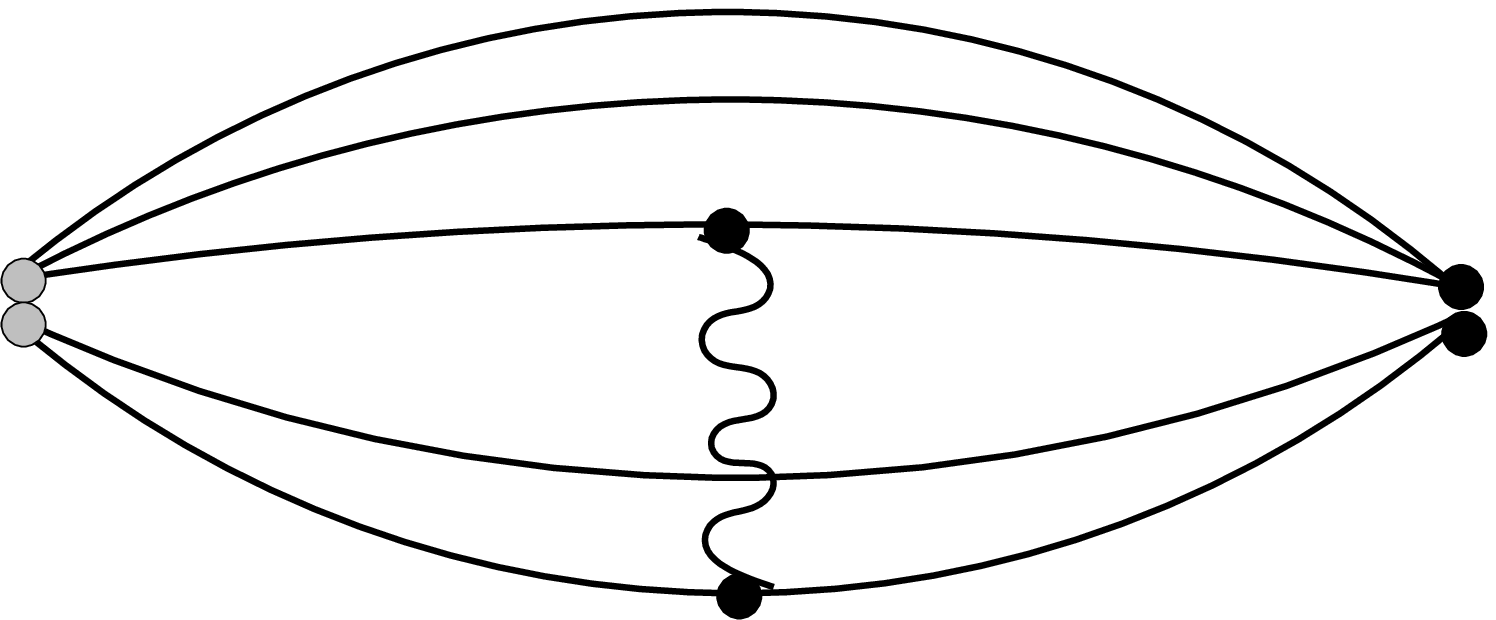}\quad
\epsfysize=1.7cm\epsffile{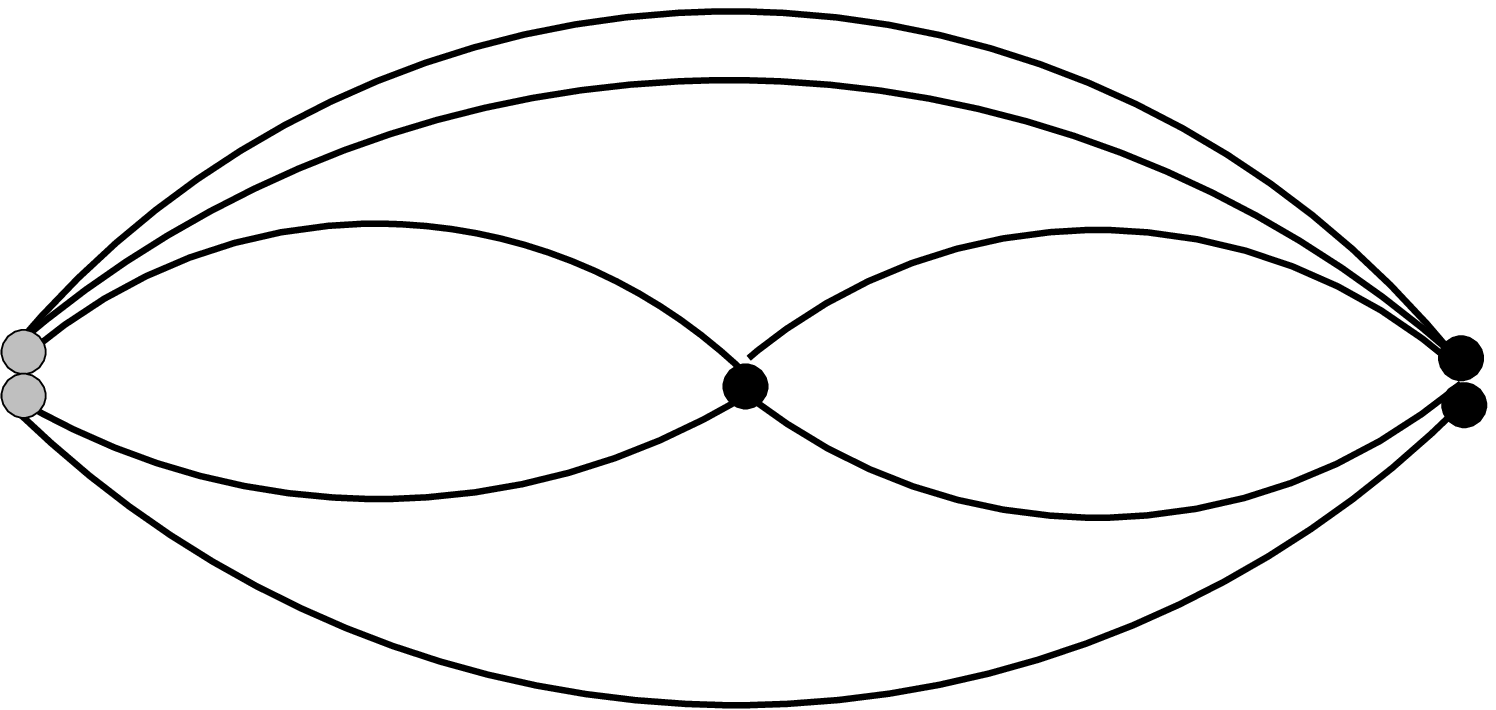}}
\caption{Feynman diagrams contributing to the order $g^2_{{\rm YM}}$ 
corrections to the correlator $\langle 0| {\rm Tr}Z^J(x)\, {\rm Tr}
\bar{Z}^J(0) |0\rangle $.
There are either $J$ scalar lines with one insertion of a scalar
self-energy sub-diagram or two lines connected by either a vector line
with two three-point vertices or one four-point scalar vertex.
\label{fig:radcorr}}
\end{figure}

The second and third of these diagrams represents the correction which
correlates four of the scalar fields.  This is the object $\Gamma$ which
we
studied in the previous section in eqs.(\ref{g})-(\ref{diagonal}).  
To take this
correction into account, we must insert the four-point module
$\Gamma$ in all appropriate ways.  The result is given by the
formula
$$
\frac{J^2}{4}\sum_{k,q=0}^{J-2}\langle 0| {\rm Tr}\left(T^aZ^k(x)T^b
Z^{J-2-k}(x)\right) {\rm Tr} \left( T^c \bar Z^q(0) T^d \bar
Z^{J-2-q}(0) \right) |0\rangle  \Gamma_{zz\bar z\bar z}^{abcd}(x)
$$ 
The factor of $J^2$ in the above expression arises from the fact that
there are $J$ places to insert the first leg of $\Gamma$ into each
of the two traces in the correlator.  
The position of the second insertion on each side
is then summed over.  The factor of 4 in the denominator arises from
the fact that all crossings have already been taken into account in
$\Gamma$ so that, to prevent double counting, the insertions should be
ordered.  It is more convenient to use the cyclic symmetry of the
trace to consider all orderings and divide by their number - thus the
factor of 4.  Using the identity in (\ref{identity1}), and the
explicit form of the four-point module in eqn.(\ref{symmetric}), we
get the formula
\begin{eqnarray}
\frac{J^2}{8}\left[ \frac{ g^2_{{\rm YM}} \Gamma(\omega-1)
}{4\pi^\omega(2-\omega)(2\omega-3) }\right]
[x^2]^{2-\omega} 2^{-J} \Delta^J(x) ~~~ ~~~~~~~ \nonumber \\
~~~~~~~\times~~ \sum_{k,q=0}^{J-2}\langle  {\rm Tr}Z^k {\rm Tr}\bar Z^q
{\rm Tr} Z^{J-2-k}\bar Z^{J-2-q} -{\rm Tr} Z^k \bar Z^q Z^{J-2-k}\bar
Z^{J-2-q}\rangle
\label{vertexcorrections}
\end{eqnarray}

In order to show that the radiative corrections cancel to order
$g_{{\rm YM}}^2$, we must demonstrate that the sum of the two contributions,
(\ref{self-energy}) and (\ref{vertexcorrections}) is zero.  This
requires an identity for the matrix model correlation functions
\bea
\lefteqn{
\langle {\rm Tr}Z^{J-1}\bar Z^{J-1}-
\frac{1}{N}{\rm Tr}Z^{J-1}{\rm Tr}\bar Z^{J-1} 
\rangle =} \\
& & \frac{1}{N}
\sum_{k,q=0}^{J-2}\langle  {\rm Tr}Z^k {\rm Tr}\bar Z^q 
{\rm Tr} Z^{J-2-k}\bar Z^{J-2-q} 
-{\rm Tr} Z^k \bar Z^q Z^{J-2-k}\bar Z^{J-2-q}\rangle
\eea
This identity can easily be shown to follow from a Schwinger-Dyson
equation of the Gaussian matrix model.  Showing that it is true is
entirely equivalent to the manipulations of Lie algebra matrices which
was outlined in section 2 of ref.\cite{D'Hoker:1998tz}.

\subsection{The first radiative correction to the anomalous dimension of 
unprotected operators to leading order \mbox{in $\frac{J^2}{N}$}
\label{leadingorder}}

Here we shall be interested in calculating the first radiative correction 
to the anomalous dimension of the unprotected operators
given in~(\ref{momentumop}).
Our calculation implies evaluating 
the first radiative correction to the following
quantity
\be
\label{pqagain}
\sum_{p,q=0}^J \langle 
{\rm Tr}\left[\phi_3(x)Z^{J-p}(x)\phi_4(x) Z^p(x)\right]
{\rm Tr}\left[\phi_3(0)\bar{Z}^{q}(0)\phi_4(0) \bar{Z}^{J-q}(0)\right]
\rangle e^{\frac{2\pi i}{J}(n_1p-n_2q)} \nonumber
\ee
The calculation is conveniently split into four parts each of which is
treated making use of our modules from before
\begin{enumerate}
\item
Corrections coming from self-energies, $C_{SE}$ 

These result from replacing one free propagator of type 
$\langle  \phi \phi \rangle$  or $\langle  Z Z \rangle$ by the relevant
one of the following two modules
\bea
\langle 0| Z^a(x)\bar{Z}^b(0)|0\rangle& =
& -\delta_{ab}\,\,\frac{g^2_{{\rm YM}} N}{8 \pi^2(2-\omega)}
\Delta(x) \nonumber\\
\langle 0| \phi_{\alpha}^a(x) \phi_{\beta}^b(0)|0\rangle& =& -
\delta_{ab}\, \delta_{\alpha \beta}\,
\frac{g^2_{{\rm YM}} N}{8 \pi^2(2-\omega)}
\Delta(x) \nonumber
\eea
\item
Corrections involving the correlator of four $Z$ fields, $C_{4Z}$ 

These are obtained by replacing a pair of $Z$-propagators with the
following module which can easily be derived from the general 
correlator~(\ref{symmetric})--(\ref{diagonal})
\be
\langle 0|  Z^a(x)Z^b(x)\bar{Z}^c(0)\bar{Z}^d(0)|0\rangle
=\frac{g^2_{{\rm YM}}}{4\pi^2(2-\omega)}
\left[f^{ead}f^{ebc}+f^{eac}f^{ebd}\right]\left(\Delta(x)\right)^2
\label{4Zmodule}
\ee
\item
Corrections involving the correlator of four $\phi$-fields, $C_{4\phi}$ 

These we get by replacing the two $\phi$-propagators with the following 
quantity
\be
\langle 0| \phi_3^a(x)\phi_4^b(x)\phi_3^c(0)\phi_4^d(0)| 0\rangle
=\frac{g^2_{{\rm YM}}}{4\pi^2(2-\omega)}
\left[f^{ead}f^{ebc}\right]\left(\Delta(x)\right)^2
\label{4pmodule}
\ee
\item
Corrections involving the correlator of two $Z$-fields and two
$\phi$-fields,
$C_{2Z\phi}$

These are obtained by replacing one $Z$-propagator and one
$\phi$-propagator
with the following module
\be
\langle 0| \phi_3^a(x)Z^b(x)\phi_3^c(0)\bar{Z}^d(0)|0\rangle
=\frac{g^2_{{\rm YM}}}{4\pi^2(2-\omega)}
\left[f^{ead}f^{ebc}\right]\left(\Delta(x)\right)^2
\label{2pZmodule}
\ee
\end{enumerate}
Inserting the given modules in all possible ways and taking care of the 
remaining contractions by use of the trace formulas in
section~\ref{notation}
we obtain to the leading order in $N$ and all orders in $J$
\bea
C^0_{SE}&=& A_{0}\cdot 
\cases{-\frac{J}{2}-1 & $n_1\neq n_2$\cr 
 -\frac{J^2}{2}-\frac{3J}{2}-1 &$n_1= n_2$\cr} \nonumber \\
C^0_{4Z}&=& A_{0}\cdot 
\cases{\frac{J}{2} &~~~~~~~~ $n_1\neq n_2$\cr 
 \frac{J^2}{2}-\frac{J}{2} &$~~~~~~~~~n_1=n_2$\cr} \nonumber \\
C^0_{4\phi}&=& A_{0}\cdot 
\cases{1&  ~~~~~~~~~~~~~~~ $n_1\neq n_2$\cr 
 1 & ~~~~~~~~~~~~~~~~$n_1= n_2$\cr} \nonumber \\
C^0_{2\phi Z}&=& A_{0}\cdot 
\cases{0 &~$n_1\neq n_2$\cr 
 2J\cos\left(\frac{2\pi n_2}{J}\right) &~$n_1= n_2$\cr} \nonumber 
\eea
where
\be
A_0=\frac{g_{{\rm YM}}^2N}{4\pi^2(2-\omega)}\ \left(\frac{1}{2}\right)^{J+2}
N^{J+2} \left[x^2\right]^{2-\omega} \Delta^{J+2}(x)
\ee
It is easy to understand the combinatorial factors of $(J+2)$ and
$(J+1)(J+2)$ 
occurring in the self-energy correction. The factor $(J+2)$ counts the
number of
propagators and the $(J+1)$ the number of momentum states.
The results neatly sum up to
\bea
C&\equiv& C_{SE}+C_{4Z}+C_{4\phi}+C_{2Z\phi} \nonumber \\
&=&\cases{0 & $n_1\neq n_2$\cr 
A_0 \cdot 2J\left(\cos\left(\frac{2\pi n_2}{J}\right)-1\right) &$n_1=
n_2$\cr} \nonumber
\eea
The contribution to the correction of the anomalous dimension coming
from planar diagrams thus reads
\be
\left(\delta \Delta\right)_0=\frac{g_{{\rm YM}}^2 Nn^2}{J^2}
\ee
which is the same as the result obtained in
reference~\cite{Berenstein:2002jq}.
We note, however, that it is crucial for the cancellations to take place
that we include $p,q=0$ in our summation range (\ref{pqagain}).
\subsection{The first radiative correction to the anomalous dimension
of unprotected operators to two orders in $\frac{J^2}{N}$ 
\label{nexttoleading}}

Calculation of the first radiative correction to the anomalous dimension 
can be carried out to higher orders in the double scaling parameter
following
exactly the same recipe as in the previous section. For the explicit
computations it is  useful to employ an effective matrix model
interaction vertex representing the four point modules of
\eqn{4Zmodule}, \eqn{4pmodule} and \eqn{2pZmodule}.
These computations are rather technical and we have chosen to
present them in detail in appendix B.

The outcome of this analysis is that up to next to leading order
in $N$ the correlation function in question at one-loop
reads
\bea
\lefteqn{
\langle\, {\cal O}_{n_1}(x)\, \bar{\cal O}_{n_2}(0)\, \rangle_{\mbox{\tiny
1-loop}} =}\nn\\&& 
-\frac{g^2_{{\rm YM}}\, N}{J^2}\, \frac{[x^2]^{2-\omega}}{4\pi^2\, (2-\omega)}
\, 
\left ( \frac{g^2_{{\rm YM}}}{8\pi^2\, |x|^2}\right )^{J+2}\,
\Bigg[\, 2 \, n_1\, n_2\, (2\pi)^2\, \langle {\cal O}_{n_1}\,\bar {\cal O}_{n_2}
\rangle_{\mbox{\tiny 0-loop}} +\frac{J^4}{N^2}\, {\cal D}_{n_1,n_2}\,
\Bigg]\nn\\&&
\eea
where
\be 
\langle {\cal O}_{n_1}\, \bar{\cal O}_{n_2}
\rangle_{\mbox{\tiny 0-loop}} = \delta_{n_1,n_2} + \frac{J^4}{N^2}
\, M_{n_1, n_2}
\ee
is the tree-level correlator of \eqn{0loop} and where ${\cal D}_{n_1,n_2}$
is the symmetric genus one mass renormalization matrix given by
($\bar n \neq 0$)
\be
{\cal D}_{0,0}=0 \qquad {\cal D}_{0,\bar n}=0
\qquad
{\cal D}_{\bar n_1, \bar n_2}=
\cases{ \frac{2}{3}+ \frac{5}{\bar n_1{}^2\, \pi^2} & 
 for $|\bar n_1|=|\bar n_2|$ \cr
 \frac{2}{3}+ \frac{2}{\bar n_1{}^2\, \pi^2}
 +\frac{2}{\bar n_2{}^2\, \pi^2} & for $|\bar n_1| \neq |\bar n_2|$} 
\ee
This result teaches us that for the redefined operators
$\tilde{\cal O}_n(x)$ of \eqn{redefinedops} we have
\bea
\lefteqn{
\langle\, \tilde{\cal O})_{n_1}(x)
\, \tilde{\bar{\cal O}}_{n_2}(0)\, \rangle_{\mbox{\tiny
1-loop}} =
-\frac{g^2_{{\rm YM}}\, N}{J^2}\, \frac{[x^2]^{2-\omega}}{4\pi^2\, (2-\omega)}
\, 
\left ( \frac{g^2_{{\rm YM}}}{8\pi^2\, |x|^2}\right )^{J+2}\,}
\nn\\&& 
\Bigl [ \,2\, n_1\, n_2\, (2\pi)^2\, \delta_{n_1,n_2} 
-(n_1-n_2)^2\, (2\pi)^2\, \frac{J^4}{N^2}\,M_{n_1, n_2}
+\frac{J^4}{N^2}\, {\cal D}_{n_1,n_2}\,
\Bigr ]\nn\\&&
\eea
The diagonal piece of this result gives rise to a
non-planar correction of the
one-loop anomalous dimension of these operators
\be
(\delta\Delta)_1=g^2_{{\rm YM}}\, \frac{J^2}{(2\pi)^2\,N^2}\,
\Bigl [ \, \frac{1}{3}+ \frac{5}{2\,\bar n{}^2\, \pi^2}\,  \Bigr ]
\, .
\ee
It would be very interesting to reproduce this result from
a genus one string theory calculation.

%%%%%%%%%%%%%%%%%%%%%%%%%%%%%%%%%%%%%%%%%%%%%%%%%%%%%%%%%%%%%%%%%%
\vspace{6mm}
\noindent
{\bf Acknowledgments} 

\noindent
We would like to thank Gleb Arutyunov and Niklas
Beisert for interesting discussions. J.P.~and M.S.~would like to thank
the University of British Columbia for hospitality during the initial
stages of this project.

\appendix

\sect{Complex matrix model technology}

Much of our insight into the nature of the scaling limit
eq.(\ref{limit}) comes from the exact result eq.(\ref{corr})
which we have not been able to find in the literature. Here we
will explain its derivation using matrix model techniques.

We start from the normalized Gaussian measure for complex 
$N\times N$ matrices $Z_{i,j}=$Re$Z_{i,j}+i~$Im$Z_{i,j}$ 
with $\bar Z_{i,j}=$Re$Z_{j,i}-i~$Im$Z_{j,i}$ 
\be
\big[ dZ d\bar{Z} \big] 
=dZ d\bar{Z}~\exp(-{\rm Tr} Z\bar{Z}),
\ee
where the flat measure $dZ d\bar{Z}$ is an abbreviation of
\be
dZ d\bar{Z}=
\prod_{i,j=1}^{N} { d{\rm Re}Z_{i,j}~d{\rm Im}Z_{i,j} \over \pi}
\ee
Thus $\int \big[ dZ d\bar{Z} \big]=1$, and matrix model expectation
values are given by 
$\langle  \ldots \rangle =\int \big[ dZ d\bar{Z} \big] \ldots $.
The model has a $U(N) \times U(N)$ symmetry (since the action
and the measure are invariant under independent left and right
group multiplication $Z \rightarrow U Z V^{\dagger}$ where $U,V$ are
unitary). Correlators invariant under the full symmetry
(i.e.~sums of products of traces of powers of $Z \bar Z$) 
are calculable with known techniques~\cite{Ambjorn:1992xu}.
Correlators with less symmetry are more difficult  
even though the potential is Gaussian. For our purposes we are 
interested in correlators invariant under the adjoint action of just
one $U(N)$ symmetry: $Z \rightarrow U Z U^{\dagger}$. 
Now the general invariant correlator contains traces of arbitrary
words made out of $Z$ and $\bar Z$, a problem that has not been solved
to our knowledge. However, in the special case where the traces
contain either just $Z$'s or $\bar Z$'s the problem is solvable
by character expansion techniques \cite{Kostov:1996bs}, 
or, alternatively, by the method of Ginibre \cite{Ginibre}.
For completeness we will briefly explain both.

In order to apply the first method one expands the parts of the
correlator containing the $Z$'s and the $\bar Z$'s separately 
in the basis of unitary group characters
(Schur functions) $\chi_h(Z)$ and $\chi_{h'}(\bar Z)$ 
labeled by Young diagrams $h$,$h'$. Then one uses
\cite{Kostov:1996bs} that, even though the matrices are complex
instead of unitary, the inner product
\be 
\langle  \chi_h(Z)~\chi_{h'}(\bar Z) \rangle=
\delta_{h,h'}~\Omega_{h}
\label{ortho} 
\ee
is still orthogonal, with a Young diagram dependent, explicitly known
normalization factor $\Omega_h$ (see \cite{Kostov:1996bs} for precise
definitions). By Schur-Weyl duality the expansion
coefficients are expressed with the help of the characters of the 
symmetric group $S_N$. E.g.~in the simplest case
$\langle {\rm Tr}Z^J~ {\rm Tr}\bar Z^J \rangle$ one has
\be
{\rm Tr}Z^J = \sum_h {\rm ch}_h(J)~\chi_h(Z)
\ee 
where ch$_h(J)$ is the character of the symmetric group $S_J$ 
corresponding to the representation $h$ and evaluated for the
conjugacy class of $J$-cycles.
Thus, using eq.(\ref{ortho}),
\be
\langle {\rm Tr}Z^J~ {\rm Tr}\bar Z^J \rangle=
\sum_h \Big( {\rm ch}_h(J) \Big)^2~\Omega_h
\ee 
Now it is easy to show that for $J-$cycles
ch$_h(J)$ is non-zero only for Young diagrams consisting of a single
hook of boxes with row length $k$ and column length $J-k$. 
For the hooks one has ch$_h(J)=\pm 1$ and (see  \cite{Kostov:1996bs})
$\Omega_h=\prod_{i=1}^k (N-1+i)~\prod_{m=1}^{J-k} (N-m)$
and thus (clearly $\sum_h$ becomes a sum over the $J$ possible hooks)
\be
\langle {\rm Tr}Z^J~ {\rm Tr}\bar Z^J \rangle=
\sum_{k=1}^J~
\prod_{i=1}^k (N-1+i)~\prod_{m=1}^{J-k} (N-m)
\ee
which may be summed to the expression eq.({\ref{corr}).  
In a similar way we find 
\be
\langle {\rm Tr} Z^J {\rm Tr} Z^K {\rm Tr}\bar{Z}^{J+K}\rangle=
\Bigg( 
\sum_{k=K+1}^{J+K} -\sum_{k=1}^J
\Bigg)~
\prod_{i=1}^k (N-1+i)~\prod_{m=1}^{J+K-k} (N-m)
\ee
which upon summing gives eq.(\ref{tpoint}).

An alternative method consists in diagonalizing 
$Z$ by a similarity transformation
\be
X Z X^{-1}=\mbox{diag}(z_1,\ldots,z_N),
\ee
with $X$ a complex matrix and $z_1,\ldots,z_N$ the complex eigenvalues of
$Z$.
As shown by Ginibre in 1965~\cite{Ginibre} the non-diagonal degrees of
freedom
can be integrated out leading to the following joint probability density
for
the eigenvalues of $Z$
\be
P(z_1,\ldots,z_N)=K^{-1} \exp\left( -\sum_{i=1}^N |z_i|^2\right)
\prod_{1\leq i\leq j\leq N}|z_i-z_j|^2,
\hspace{0.5cm}K= \pi^N\prod_{j=1}^N j! 
\ee
Here the normalization is such that
\be
\int_{C \hspace{-0.18cm}|}\, \prod _{i=1}^N  d^{\,2} z_i\,
P(z_1,\ldots,z_N)=1.
\ee
Furthermore, one can derive an explicit expression for the correlation
function involving $n$ eigenvalues, defined as follows $(n< N)$
\be
R_n(z_1,\ldots,z_n)\equiv  \int_{C\hspace{-0.18cm}|}\,
\prod_{i=n+1}^N  d^{\,2}z_i \,P(z_1,\ldots,z_N).
\ee
The result reads~\cite{Ginibre}
\be
R_n(z_1,\ldots,z_n)=\pi^{-n}\, \frac{(N-n)!}{N!} 
\exp\left(-\sum_{k=1}^n |z_k|^2\right) 
{\mbox{det}}(K_N(z_i,z_j))|_{i,j=1,\ldots,n}
\label{Rn}
\ee
where
\be
K_N(z_i,z_j)=\sum_{l=0}^{N-1}\frac{(z_i z_j^*)^l}{l!}.
\ee
Making use of the expression~(\ref{Rn}) for $n=1$, $n=2$ and $n=3$ 
one easily derives~(\ref{corr}) and~(\ref{tpoint}).

\sect{Evaluating $C_{4Z}$,$C_{2\phi Z}$,$C_{4\phi}$ 
and $C_{SE}$}

\subsection{$C_{4Z}$}

Let us begin with the computation of  $C_{4Z}$, i.e.\ the insertion 
of the module \eqn{4Zmodule} into the correlator. Here it is useful
to represent the expression \eqn{4Zmodule} through an effective
four point vertex in the matrix model of the form $:\Tr(Z^2\Zb^2-Z\Zb
Z\Zb):$.
This effective vertex appears in a normal ordered fashion, disallowing
self-contractions. After performing the $\phi_3$ and $\phi_4$
contractions one is then led to evaluate the correlator
\be
\langle \Tr Z^{J-p}\,\Zb^{J-q} \, \Tr Z^p\, \Zb^q\, :\Tr(Z^2\Zb^2-Z\Zb
Z\Zb):
\rangle
\nn
\ee
Contacting the two $Z$'s in the effective vertex with the outside
$\Zb$'s one produces, after some manipulations, the following three terms 
\bea
(A)&=&\sum_{l=0}^{q-1}\, \sum_{m=0}^{J-q-1}\, \langle 
\Tr( Z^{J-p}\,\Zb^{m+q+1-l}\, Z^p\, \bar{Z}^{J+l-q-1-m} )
\nn\\&&\qquad\qquad
- \Tr (Z^{J-p}\,\Zb^{m+q-l}\, Z^p\,
\bar{Z}^{J+l-q-m}) 
\rangle 
\nn\\
(B)&=&\sum_{l=1}^{q-1}\, \sum_{m=0}^{l-1}\, \langle \Tr(Z^{J-p}\,
\Zb^{J-q})\,
\Bigl( \Tr\Zb^m\, \Tr(Z^p\,\Zb^{q-m}) -
\Tr\Zb^{m+1}\, \Tr(Z^p\,\Zb^{q-m-1})\, \Bigr )\rangle\nn\\
(C)&=&\sum_{l=0}^{q-2}\, \sum_{m=0}^{q-l-2}\, \langle \Tr(Z^{J-p}\,
\Zb^{J-q})\,
\Bigl( \Tr\Zb^{m+2}\, \Tr(Z^p\,\Zb^{q-m-2})\nn\\&&\qquad\qquad -
\Tr\Zb^{m+1}\, \Tr(Z^p\,\Zb^{q-m-1})\, \Bigr )\rangle\nn
\eea
to be augmented by the corresponding terms obtained by swapping
$p\rightarrow J-p$ and $q\rightarrow J-q$. We are interested in these
correlators to orders $N^{J+3}$ and $N^{J+1}$. Term (A) is of
leading order $N^{J+1}$ and hence its evaluation is simple:
swap $m\rightarrow J-q-1-m$ in the second sum to obtain
\bea
(A)&=& \sum_{l=0}^{q-1}\, \sum_{m=0}^{J-q-1}\, \langle 
\Tr( Z^{J-p}\,\Zb^{J-m-l}\, Z^p\, \bar{Z}^{l+m} ) 
- \Tr( Z^{J-p}\,\Zb^{J-m-l-1}\, Z^p\,
\bar{Z}^{l+m+1})\rangle 
\nn\\
&=& \sum_{l=0}^{q-1}\, \langle \Tr(Z^{J-p}\, \Zb^{J-l}\, Z^p\, \Zb^l)\,
\rangle - \sum_{l=0}^{q-1}\, \langle \Tr(Z^{J-p}\, \Zb^{l+1}\, Z^p\, 
\Zb^{J-1-l})\,\rangle\nn\\
&=& N^{J+1}\, \sum_{l=0}^{q-1}\, \mbox{Min}[J-p,J-l,p,l] -
\sum_{l=1}^{q}\, \mbox{Min}[J-p,l,p,J-l] 
\nn\\
&=& - N^{J+1}\,
\mbox{Min}[J-p,J-q,p,q]
\eea
up to terms of order $N^{J-1}$.  Turning to
the contributions (B) and (C) we observe that they also are of 
telescope type and only the lowest and highest values of the 
summation indices $m$ contribute:
\bea
(B)&=& \sum_{l=1}^{q-1}\langle\, \Tr(Z^{J-p}\, \Zb^{J-q})\,
\Bigl( N\, \Tr(Z^p\,\Zb^{q}) -
\Tr\Zb^{l}\, \Tr(Z^p\,\Zb^{q-l})\, \Bigr )\rangle\nn\\
   &=& (q-1)\, N\, \langle\, \Tr(Z^{J-p}\, \Zb^{J-q})\,\Tr(Z^p\,\Zb^{q})\,
\rangle\nn\\
&&\quad -\sum_{l=1}^{q-1}\, \langle\, \Tr(Z^{J-p}\, \Zb^{J-q})\,
\Tr\Zb^{l}\, \Tr(Z^p\,\Zb^{q-l})\, \rangle\nn\\
(C)&=& \sum_{l=0}^{q-2}\langle \Tr(Z^{J-p}\, \Zb^{J-q})\,
\Bigl( \Tr\Zb^{q-l}\, \Tr(Z^p\,\Zb^{l}) -
\Tr\Zb\, \Tr(Z^p\,\Zb^{q-1})\, \Bigr )\nn\\
&=& \sum_{l=2}^q\, \langle \Tr\,( Z^{J-p}\, \Zb^{J-q})\,\Tr \Zb^l
\, \Tr (Z^p\,\Zb^{q-l})\,\rangle \nn\\&&\quad -(q-1)\,
\langle \Tr\,( Z^{J-p}\, \Zb^{J-q})\Tr\Zb\, \Tr(Z^p\,\Zb^{q-1})\,
\rangle
\eea
But now the second term in (B) telescopes with the first term in
(C) and we obtain 
\bea
\lefteqn{
(B)+(C)=}\nn\\
&&(q-1)\, N\, \langle\, \Tr (Z^{J-p}\,\Zb^{J-q})\, \Tr (Z^p\, \Zb^q)\, 
\rangle
-q\, \langle\, \Tr\Zb\, \Tr Z^p\, \Zb^{q-1}\, \Tr
Z^{J-p}\,\Zb^{J-q}\rangle\nn\\&& \quad
 + \langle\, \Tr (Z^{J-p}\,\Zb^{J-q})\, \Tr\Zb^q\, \Tr Z^p\,
\rangle
\eea
Adding in the swapped $p\rightarrow J-p$ and $q\rightarrow J-q$ 
contributions we thus find the result
\bea
C^1_{4Z}&=&(J-2)\, N\, (J|\, p,q)- 2\, N^{J+1}\, \mbox{Min}[J-p,J-q,p,q]
\nn\\&&
+\langle\, \Tr (Z^{J-p}\,\Zb^{J-q})\, \Tr\Zb^q\, \Tr Z^p\,
\rangle
+\langle\, \Tr (Z^{p}\,\Zb^{q})\, \Tr\Zb^{J-q}\, \Tr Z^{J-p}\,
\rangle
\nn\\&&
-q\, p\, (J-1|\, p-1,q-1) - q\, (J-p)\, (J-1|\, p,q-1)
\nn\\&&
-(J-q)\, p\, (J-1|\, p-1,q) - (J-q)\, (J-p)\, (J-1|\, p,q)
\eea
up to terms of order $N^{J-1}$ and where we have defined
\be
(J|\, p,q) \equiv \langle
\Tr ( Z^{J-p}\, \Zb^{J-q}\,)\, \Tr(Z^p\,\Zb^q\,) \rangle
= \delta_{p,q}\,[J|\, p]+  [J|\, p,q] \, .
\ee

\subsection{$C_{2\phi Z}$}

For the evaluation of $C_{2\phi Z}$ we use an analogous strategy
and represent the insertion of the module \eqn{2pZmodule} into
\eqn{pqagain} by an effective matrix model vertex of the
form
\be
V_{2\phi Z}\equiv \frac{g^2_{{\rm YM}}\, 
\Delta(x)^2}{4\pi^2\, (2-\omega)}\, \frac{1}{8}\,
:\, \Tr\, [\phi,\Zb]\, [\bar\phi,Z]\, :
\label{V4}
\ee
where we have introduced the complex matrices $\phi$ and $\bar\phi$
corresponding to the {\it real} fields $\phi_i(x)$ and $\phi_i(0)$ with
$i=3,4$. Note, that one needs to work with complex matrices in the effective
matrix model description of the combinatorics 
for the real fields $\phi_i(x)$ in order to ensure
that only fields at points $x$ and $0$ are contracted and not at $0-0$
or $x-x$. This is a purely technical maneuver to ensure correct
combinatorics. The $:\,\,\, :$ denote normal 
ordering as before. It is easy to convince oneself that \eqn{V4} is the
correct object by inserting it into a trial correlator
\bea
\lefteqn{
\langle \Tr (\phi\, A\, Z \, B)\, \Tr (\bar\phi \, C\, \Zb D)\,V_{2\phi Z}
\rangle =}\nn\\&&
\Tr A\, \Tr C\, \Tr BD + \Tr B\, \Tr D\, \Tr AC - \Tr DACB-
 \Tr ADBC
\eea
as it should be from (\ref{identity1}) and  (\ref{2pZmodule}).

The $C_{2\phi Z}$ correlator then reads
\be
C_{2\phi Z}= \langle \Tr(\phi\, Z^{J-p}\, \phi_4\, Z^p)\,
\Tr(\bar\phi\, \Zb^q\, \bar\phi_4\, \Zb^{J-q})\,
:\Tr(\phi\Zb\bar\phi Z + \Zb\phi Z \bar \phi - \Zb\phi \bar\phi Z
-\phi \Zb Z\bar\phi):\rangle
\ee
where we have chosen to insert a $2\phi_3\, Z$ module. Clearly for the second
choice $2\phi_4\, Z$ one gets a factor of 2.
Now contract the $\bar\phi$ in the effective vertex to  find
\be
\langle\,
\Tr\Bigl(  Z^{J-p}\, \phi_4\, Z^p\, :(Z\phi\Zb + \Zb\phi Z 
- Z \Zb\phi 
-\phi \Zb Z ):\Bigr )\, \Tr(\bar\phi\, \Zb^q\, \phi_4\, \Zb^{J-q}\,) \rangle
\ee
now contract the $\phi_4$'s
\be
\langle\,
\Tr(  Z^{J-p}\, \Zb^{J-q}\, \bar\phi\, \Zb^q\,
 Z^p\, :(Z\phi\Zb + \Zb\phi Z - Z \Zb\phi 
-\phi \Zb Z ):\,\rangle
\ee
and finally contract the remaining $\phi$'s. In order to keep
track of the normal ordered $Z$ and $\Zb$ in the effective vertex we
denote them by calligraphic letters ${\cal Z}$ and $\bar{\cal Z}$.
One then has the four terms
\bea
&\langle\,
\Tr ( Z^{J-p}\, \Zb^{J-q}\, \bar{\cal Z})\, \Tr(\Zb^q\, Z^p\,{\cal Z})
+ \Tr ({\cal Z}\, Z^{J-p}\, \Zb^{J-q}\, )\, \Tr(\bar{\cal Z}\, \Zb^q\, Z^p\,)
&\nn\\&
-\Tr ( Z^{J-p}\, \Zb^{J-q}\, )\, \Tr(\bar{\cal Z}\, \Zb^q\, Z^p\,{\cal Z})
-\Tr ( {\cal Z}\, Z^{J-p}\, \Zb^{J-q}\,\bar{\cal Z} )\, \Tr (\Zb^q\, Z^p)
\,\rangle &
\eea
In the above we are not allowed to contract the ${\cal Z}$ and $\bar{\cal Z}$
with each other. However, we may circumvent this problem by forgetting
about this property of the ${\cal Z}$'s in the propagator and subtracting off
the ${\cal Z}\bar {\cal Z}$ contraction to correct our mistake, e.g.
\bea
\lefteqn{
\langle
\Tr ( Z^{J-p}\, \Zb^{J-q}\, \bar{\cal Z})\, \Tr(\Zb^q\, Z^p\,{\cal Z})
\rangle
= }\nn\\&&\langle
\Tr ( Z^{J-p}\, \Zb^{J+1-q}\,)\, \Tr(\Zb^q\, Z^{p+1}\,) \rangle - 
\langle \Tr(Z^J\,\Zb^J)
\rangle
\eea
But now the correlator is easy to evaluate and one finds
\bea
C_{2\phi Z} &=& (J+1|\, p,q+1) + (J+1|\, p+1,q)\nn\\&&
 -(J+1|\, p+1,q+1)
-(J+1|\, p,q)\nn\\
&& + 2\, N\, (J|\, p,q) - 2\, \langle \Tr(Z^J\,\Zb^J)
\rangle
\eea
which is the exact result. The total contribution to the
correlator comes with a factor of 2 and we know every term in 
the above to leading and subleading order in $N$. 

\subsection{$C_{4\phi}$}

To deduce the $C_{4\phi}$ contribution one starts out from
\be
\langle \Tr(T^a\, Z^{J-q}\, T^b\, Z^q)\, \Tr(T^c\, \Zb^p\, T^d
\, \Zb^{J-p})\, \rangle \, \langle 0| \phi_3^a\, \phi_4^b\,
\phi_3^c\,\phi_4^d\, |0\rangle
\ee
and inserts the $C_{4\phi}$ module \eqn{4pmodule} into the last term. Upon
using the identity \eqn{identity1} one finds
\bea
C_{4\phi} &=& \langle\, \Tr Z^{J-q}\, \Tr\Zb^p\, \Tr (Z^q\, \Zb^{J-p})
\rangle
+  \langle\Tr Z^q\, \Tr\Zb^{J-p}\, \Tr (Z^{J-q}\, \Zb^p)
\rangle\nn\\&&
 -\langle\Tr(Z^{J-q}\, \Zb^p\, Z^q\, \Zb^{J-p})\rangle-
 \langle\Tr(Z^{J-q}\, \Zb^{J-p}\, Z^q\, \Zb^{p})\rangle\nn\\
&=&  \langle\, \Tr Z^{J-q}\, \Tr\Zb^p\, \Tr (Z^q\, \Zb^{J-p})
\rangle
+  \langle\Tr Z^q\, \Tr\Zb^{J-p}\, \Tr (Z^{J-q}\, \Zb^p)
\rangle\nn\\
&&- 2\, N^{J+1}\, \mbox{Min}[J-p,J-q,p,q]
\eea
where we have used that $\langle\Tr(Z^{J-q}\, \Zb^p\, Z^q\, \Zb^{J-p})\rangle
= N^{J+1}\, \mbox{Min}[J-p,J-q,p,q]$ to leading order in $N$.

\subsection{$C_{SE}$}

We now turn to the contribution from the self-energy sector. It is
simply given by $-N$ times the number of free propagators
of the tree-level result plus a correction piece due to 
the missing $U(1)$ contractions in the 1-loop corrected
propagator due to \eqn{identity2}. The outcome is
\bea
C_{SE}&=& -(J+2)\, N\, (J|p,q) + 2\, \Tr (Z^J\, \Zb^J)\nn\\
&& + p(J-q)\, (J-1|p-1,q) + p\, q\, (J-1|p-1,q-1)\nn\\
&& + (J-p)\, (J-q)\, (J-1|p,q) + (J-p)\, q\, (J-1|p,q-1) \, .
\eea

\subsection{The Sum and its Fourier Transform}

In summing up all the results of subsections B.1-4
 one sees that many terms cancel out. We are then left with
\bea
\lefteqn{2\, C_{2\phi Z} + C_{4\phi} + C_{4 Z} + C_{SE} =}\nn\\
&&
2\,\Bigl\{ \,(J+1|\, p,q+1) + (J+1|\, p+1,q) 
 -(J+1|\, p+1,q+1)
-(J+1|\, p,q)\, \Bigr \}\nn\\
&& +\langle\, \Tr (Z^{J-p}\,\Zb^{J-q})\, \Tr\Zb^q\, \Tr Z^p\,
\rangle
+\langle\, \Tr (Z^{p}\,\Zb^{q})\, \Tr\Zb^{J-q}\, \Tr Z^{J-p}\,
\rangle \nn\\
&&+\langle\, \Tr Z^{J-q}\, \Tr\Zb^p\, \Tr (Z^q\, \Zb^{J-p})
\rangle
+  \langle\Tr Z^q\, \Tr\Zb^{J-p}\, \Tr (Z^{J-q}\, \Zb^p)
\rangle\nn\\
&& -4\langle \Tr Z^J\Zb^J\rangle -4 \,   N^{J+1}\, \mbox{Min}[J-p,J-q,p,q]
\label{Resultat}
\eea
In the above all terms except $\langle \Tr Z^J\Zb^J\rangle$ will turn
out to be relevant in the double scaling limit \eqn{limit}.

The first line of the right hand side of \eqn{Resultat} may be 
Fourier transformed by multiple telescoping. One finds
\bea
\lefteqn{
2\,\sum_{p,q=0}^J\Bigl\{ \,(J+1|\, p,q+1) 
\ldots -(J+1|\, p,q)\, \Bigr \}\, z^p\, w^q=}\nn\\
&&
-2\, (1-z^{-1})\, (1-w^{-1})\, \sum_{p,q=1}^J\, (J+1|\, p,q)\, z^p\, w^q
\nn\\&&
-4\, N\, \langle \Tr Z^{J+1}\, \Zb^{J+1}\rangle
+4\, \langle \Tr Z^{J+1}\, \Tr \Zb^{J+1}\rangle
+8 J\, N^{J+1}
\label{firstres}
\eea
where $z=\exp(2\pi\, i\, n_1/J)$ and $w=\exp(-2\pi\, i\, n_2/J)$.
We thus recover the genus-0 correlator $\langle {\cal O}_{n_1}\,
\bar {\cal O}_{n2}\rangle_{\mbox{\tiny 0-loop}}$ in the sum of
the second line\footnote{The shift of $J\rightarrow J+1$ in this 
term is irrelevant in the scaling limit and results in an additional
factor of $N$ in the tree-level correlator.}. Note that there 
is a relevant contribution to the scaling limit from the first
term in the last line of (\ref{firstres}).

To evaluate the minimum sum 
\be
-4\,\sum_{p,q=0}^J\, \mbox{Min}[J-p,p,J-q,q]\, z^p\, w^q
\label{min1}
\ee
we proceed as follows. Split up the individual $p$ and $q$ sums
according to $\sum_{p=0}^J=\sum_{p=0}^{J/2}+\sum_{p=J/2+1}^J$
assuming $J$ even. By reversing the orders of the resulting sums over
$p$ and $q$ in the domain larger than $J/2$ via $p'=J-p$ and
$q'=J-q$ one can turn \eqn{min1} into
\be
-4\,\sum_{p,q=0}^{J/2}\, \mbox{Min}[p,q]\, (z^q+z^{-q})\, (w^q+w^{-q})
- (\mbox{subleading terms of order} J^2) 
\label{min2}
\ee
Now this apparently is
\be
{\cal A}_{n_1,n_2}=
-16\,\sum_{p=0}^{J/2}\, \Bigl\{ \sum_{q=0}^p\, q\, \cos(\ft{2\pi\,n_1}{J} p)
\, \cos(\ft{2\pi\,n_2}{J} q)
+\sum_{q=p+1}^{J/2}\, p\, \cos(\ft{2\pi\,n_1}{J} p)
\, \cos(\ft{2\pi\,n_2}{J} q)\, \Bigr \}
\ee
which upon performing the sums explicitly turns out to be
$$
{\cal A}_{0,0}=-\frac{2\,J^3}{3}
\qquad {\cal A}_{\bar n_1,0}=\frac{2\, J^3}{\bar n_1{}^2\,\pi^2}\qquad 
{\cal A}_{0,\bar n_2}=\frac{2\, J^3}{\bar n_2{}^2\,\pi^2}\qquad 
{\cal A}_{\bar n_1,\bar n_2}= 0 
$$
\be
{\cal A}_{\bar n_1,\bar n_1} = - \frac{J^3}{\bar n_1{}^2\, \pi^2}
=  {\cal A}_{\bar n_1,-\bar n_1} \qquad
\mbox{where} \quad |\bar n_1|\neq |\bar n_2|, 
\quad \bar n_1\neq 0 \neq \bar n_2
\label{minres}
\ee
What remains to be done is the Fourier transform of the
four cubic trace correlators in \eqn{Resultat} which may be rewritten
as
\be
\langle \Tr^3\rangle = \sum_{p,q=0}^J\, 
\langle \Tr Z^p\, \Tr \Zb^q\, \Tr ( Z^{J-p}\, \Zb^{J-q})\, \rangle
\, (z^p+z^{-p})\, (w^q+w^{-q})
\ee
Now
\bea
\lefteqn{\langle \Tr Z^p\, \Tr \Zb^q\, \Tr ( Z^{J-p}\, \Zb^{J-q})\, \rangle
=}\nn\\&& N^2\,\delta_{p,0}\, \delta_{q,0}\, \langle\, \Tr Z^J\, \Zb^J\,
\rangle
+\delta_{p,q}\, \langle\, \Tr Z^{J-p}\,\Zb^{J-p}\rangle
\, \langle\Tr Z^p\, \Tr \Zb^p\rangle_C
\nn\\&&
+\delta_{p,0}\, N\, \langle\Tr \Zb^q\,\Tr (Z^J\Zb^{J-q})\rangle_C
+\delta_{q,0}\, N\, \langle\Tr Z^p\,\Tr (Z^{J-p}\Zb^{J})\rangle_C
\eea
which upon using 
$$\langle \Tr Z^p\,\Tr (Z^{J-p}\Zb^{J})\rangle_C=N^{J+1}(J-p+1)\,p
+ {\cal O}(N^{J-1})$$
yields to the order in $N$ and $J,p,q$ we are working at
\bea
\lefteqn{\langle \Tr Z^p\, \Tr \Zb^q\, \Tr ( Z^{J-p}\, \Zb^{J-q})\, \rangle
=}\\
&&\delta_{p,0}\, \delta_{q,0}\, (N^{J+3}+\binomial{J+1}{4}
\, N^{J+1})
+\delta_{p,0}\, N^{J+1}\, (J-q)\, q
+\delta_{q,0}\, N^{J+1}\, (J-p)\, p \nn
\eea
Fourier transforming this result by making use of
$$
\int_0^1 dx\, x\, (1-x)\, \cos(2\pi\, n\, x) =
\cases{-\frac{1}{2\, n^2\, \pi^2} & for $n\neq 0$\cr
\quad \frac 1 6 & for $n=0$ \cr}
$$
yields
\be
\langle \Tr^3\rangle = 
4 \, \left(N^{J+3}+\binomial{J+1}{4}\, N^{J+1} \right) + 
J^3\, N^{J+1}\, 
\cases{\qquad \frac 4 3 & $n_1=0=n_2$\cr \frac{2}{3} - \frac{2}{n_2{}^2\, \pi^2}
& $n_1=0$; $n_2\neq 0$ \cr
\frac{2}{3} - \frac{2}{n_1{}^2\, \pi^2}
& $n_2=0$; $n_1\neq 0$ \cr
-\frac{2}{n_1{}^2\, \pi^2} -\frac{2}{n_2{}^2\, \pi^2} & 
$n_1\neq 0 \neq n_2$\cr}
\label{tr3res}
\ee
Finally turning to the relevant term in \eqn{firstres} we get
\be
-4N\, \langle \Tr Z^{J+1}\, \Zb^{J+1}\, \rangle
= -4\, \left(N^{J+3}+\binomial{J+2}{4}\, N^{J+1}\right )
\label{relevant}
\ee
Adding \eqn{minres}, \eqn{tr3res} and \eqn{relevant} we see that 
we obtain the final result of (working with the normalization
of the operators ${\cal O}_n$ defined in \eqn{momentumop})
\be
\langle {\cal O}_{n_1}\, \bar{\cal O}_{n_2}\rangle_{\mbox{\tiny 1-loop}}
= \Bigl \{-2\,(2\pi)^2\, n_1\,n_2\, \frac{N}{J^2}\, 
\langle {\cal O}_{n_1}\, \bar{\cal O}_{n_2}\rangle_{\mbox{\tiny 0-loop}}
- \frac{J^2}{N}\, {\cal D}_{n_1,n_2} 
\Bigr \}\,\times [\mbox{pole}]
\label{final}
\ee
where 
$$
{\cal D}_{0,0}=0 \qquad {\cal D}_{0,\bar n_2}=0= {\cal D}_{\bar n_1,0}\qquad
{\cal D}_{\bar n_1, \bar n_1}= \frac{2}{3}+ \frac{5}{\bar n_1{}^2\, \pi^2}
={\cal D}_{\bar n_1, -\bar n_1}
$$
\be
{\cal D}_{\bar n_1, \bar n_2}=  \frac{2}{3}+ \frac{2}{\bar n_1{}^2\, \pi^2}
+ \frac{2}{\bar n_2{}^2\, \pi^2}
\qquad \mbox{where}\quad |\bar n_1| \neq |\bar n_2| \,.
\ee
Indeed this correlator vanishes for $n_1=0$ or $n_2=0$.

%%%%%%%%%%%%%%%%%%%%%%%%%%%%%%%%%%%%%%%%%%%%%%%%%%%%%%%%%%%%%%%%% 

\end{document}